\documentclass[fleqn,usenatbib]{mnras}
\usepackage{times}
\usepackage{url}
\usepackage{epsfig}
\usepackage[flushleft]{threeparttable}
\usepackage{amssymb}
\usepackage{amsmath}
\usepackage{pdfpages}
\usepackage{todonotes}
\usepackage{array}
\usepackage{comment}

\hypersetup{draft}

\newcommand{\refbf}{}

\title[Spatially Dependent Photometric Activity]{Spatially Dependent Photometric Activity of M dwarfs in the Solar Cylinder}

\author[Chang, Wolf, \& Onken]{Seo-Won Chang$^{1,2,3,4}$, Christian Wolf$^{3,4}$, Christopher A. Onken$^{3,4}$ \\
$^{1}$SNU Astronomy Research Center, Seoul National University (SNU), 1 Gwanak-rho, Gwanak-gu, Seoul 08826, Korea\\
$^{2}$Astronomy Program, Dept. of Physics \& Astronomy, SNU, 1 Gwanak-rho, Gwanak-gu, Seoul 08826, Korea\\
$^3$Research School of Astronomy and Astrophysics, Australian National University, Weston Creek ACT 2611, Australia \\
$^4$Centre for Gravitational Astrophysics 
(CGA), 
Australian National University, Acton ACT 2601, Australia
\\
}

\begin{document}

\date{draft \today}

\maketitle

\begin{abstract}
We study the relationship between Galactic location ($R, Z$) and photometric activity for 3.6 million M dwarf stars within 1~kpc of the Sun. For this purpose, we identify 906 unique flare events as a proxy for magnetic activity from the SkyMapper Southern Survey DR3. We adopt vertical distance $|Z|$ from the Galactic disc as a proxy for age and confirm a strong trend of flaring fraction decreasing with growing stellar age. Among M dwarfs within 50~pc of the Sun, we find a flaring fraction of 1-in-1,500, independent of spectral type from M2 to M7, suggesting that these stars are all in a flare-saturated young evolutionary stage. We find {\refbf a hint of a kink} 
in the slope of the overall flare fraction near 100~pc from the plane, where a steep decline begins; this slope change is visible for mid-type M dwarfs (M3--M5), suggesting it is not an artefact of mixing spectral type. Together with SDSS H$\alpha$ emission, this trend is additional evidence that the activity fraction of M dwarfs depends on Galactic height and activity lifetime. While there is a hint of flattening of the overall activity fraction above $|Z|\approx$ 500~pc, our data do not constrain this further. Within $\sim$500~pc distance from the Sun, we find no sign of radial disk gradients in flare activity, which may only be revealed by samples covering a larger radial range. 
\end{abstract}

\begin{keywords}
stars: flare -- stars: low-mass --  stars: activity -- stars: statistics -- techniques: photometric
\end{keywords}

\section{Introduction}\label{sec:intro}
M dwarfs, the most common and longest-lived stars in our Galaxy, have frequent energetic outbursts. Their powerful flares emit across the entire electromagnetic spectrum and pose danger to potential life on planets orbiting them (e.g., \citealt{Vida2017ApJ...841..124V, Howard2018ApJ...860L..30H, MacGregor2021ApJ...911L..25M}). Growing attention towards high-energy radiation and its impact on habitability (e.g., \citealt{Tilley2019AsBio..19...64T, Howard2020ApJ...902..115H}) will benefit from understanding the properties of M dwarf ensembles at a variety of evolutionary stages as well as individual host stars with diverse properties. 

We know that the strength of M dwarf activity decreases as the stars get older, as this is revealed by activity tracers such as H$\alpha$ and white-light (near UV + optical continuum) flare emissions. The presence of H$\alpha$ emission is a useful diagnostic to classify stars as active or inactive (e.g., \citealt{West2004AJ....128..426W, Newton2017ApJ...834...85N}). We have clear evidence that the H$\alpha$ activity fraction of M dwarfs decreases with increasing height $|Z|$ above the Galactic plane \citep{West2006AJ....132.2507W, West2008AJ....135..785W,West2011AJ....141...97W} and depends on location in the Galaxy \citep{Pineda2013AJ....146...50P}. This is a natural consequence of an age-activity relation, which is also supported by kinematic evidence of increasing velocity dispersion of stars and their scale heights with age in the Milky Way disc \citep{Haywood2013A&A...560A.109H, Sharma2021MNRAS.506.1761S}. \citet{Kowalski2009AJ....138..633K} studied M dwarf flares in optical wavelengths and found similar evidence of strongly age-dependent flaring activity.  \citet{Chang2020MNRAS.491...39C} confirmed the trend of decreasing flaring fraction with $|Z|$ more clearly than before, suggesting also a possible flattening towards larger distances. 

Within the context of stellar evolution, activity lifetime plays an important part in shaping the age-activity relation of M dwarfs. As a consequence of weakening magnetic field strength together with rotational spin down, we expect an evolutionary transition phase from young and active to old and inactive on timescales that depend on mass \citep{West2011AJ....141...97W}; early-type M dwarfs ($<$~M3) experience more rapid decrease in activity than fully convective M dwarfs. Thanks to well-determined ages, members of star clusters are ideal targets to test mass-dependent age-activity relations. \citet{Ilin2019A&A...622A.133I, Ilin2021A&A...645A..42I} found that flare frequency declines with age from the Pleiades (age $\sim$ 135~Myr) to Ruprecht 147 (age $\sim$ 2.6~Gyr), and that the decrease is more rapid for higher mass stars. 

For individual field stars, age estimation is still highly uncertain but stellar rotation periods can be used as an appropriate proxy for age (gyrochronology: \citealt{Barnes2003ApJ...586..464B, Barnes2007ApJ...669.1167B}). With the help of Kepler field stars with measured rotation periods, \citet{Davenport2019ApJ...871..241D} found decreasing overall flare rates with increasing rotation-based age for relatively young G-M stars (mainly younger than 1~Gyr). However, there is still a dearth of old field M stars with well-constrained age (a few Gyr or more) for testing the age-activity relation in a mass-dependent manner. 


In this study we use a statistical approach to get reliable mean ensemble properties (age and activity fraction) for an M dwarf sample rather than individual stars. \citet{Chang2020MNRAS.491...39C} {\refbf completed} a similar study with a smaller M dwarf sample based on SkyMapper DR1.1. In Section~2 we describe the SkyMapper DR3 multi-colour time-series data and the resulting, extended M dwarf sample. Section~3 outlines how we select M dwarf flares with the flare variability index. The results of our age-activity relation are presented in Section~4, and we conclude in Section~5.

\section{Data}
\label{sec:data}
The empirical light-curve template of M dwarf flares has a fast-rise-exponential-decay pattern, except for the cases where flares have complex morphologies \citep{Davenport2014ApJ...797..122D}. The rise time of flare light-curves is only a few minutes, but then it may last up to a few hours before returning to the pre-event level. Due to the unpredictable nature of flare events and the limitations of ground-based observations, it has been challenging to have statistically meaningful sample of M dwarf flares with detailed light curves. One way for obtaining well-sampled data for flares is simultaneous high-cadence photometric and spectroscopic monitoring of a few stars, which is an excessive effort as one needs to keep observing until the flares happen (e.g., \citealt{Kowalski2013ApJS..207...15K,Kowalski2016ApJ...820...95K,Kowalski2019ApJ...871..167K}). Another alternative would be long-term, high-cadence monitoring of numerous M dwarfs located in the same patch of the sky (e.g.,  \citealt{Chang2015ApJ...814...35C,Howard2019ApJ...881....9H}). By using repeated snapshot  observations of the survey field, several time-domain surveys have recorded rapidly changing spectral energy distributions (SEDs) as a signature of M dwarf flares (e.g., \citealt{Kowalski2009AJ....138..633K,Davenport2012ApJ...748...58D,Osten2012ApJ...754....4O,Chang2020MNRAS.491...39C}). Although the measurements of these flare SEDs are limited by the temporal resolution, it still allows to investigate how often they occur and how much energy do they produce.

\subsection{SkyMapper DR3 light curves}
\label{subsec:DR3 lightcurves}

\begin{figure}
\includegraphics[width=\linewidth]{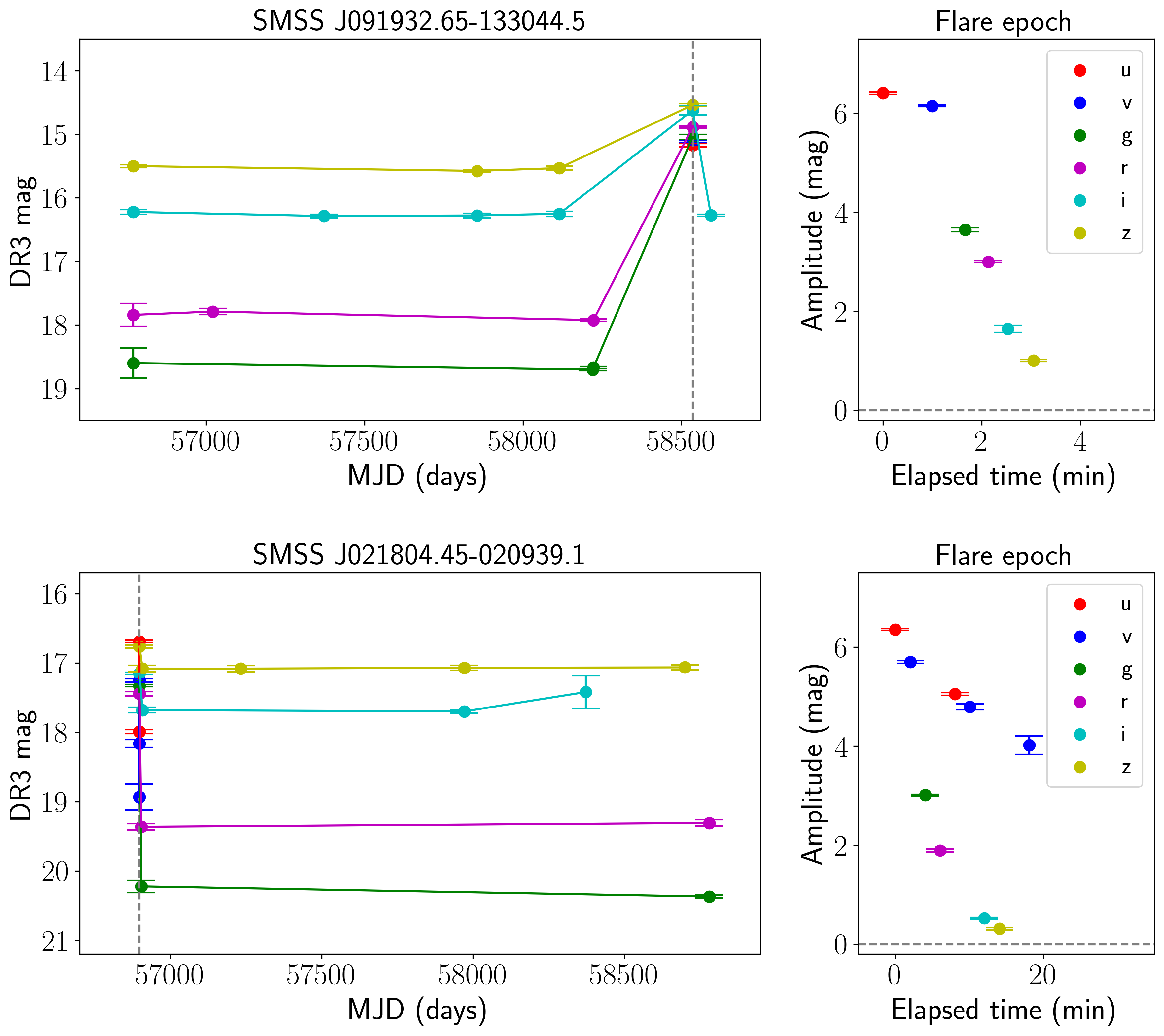}
\includegraphics[width=\linewidth]{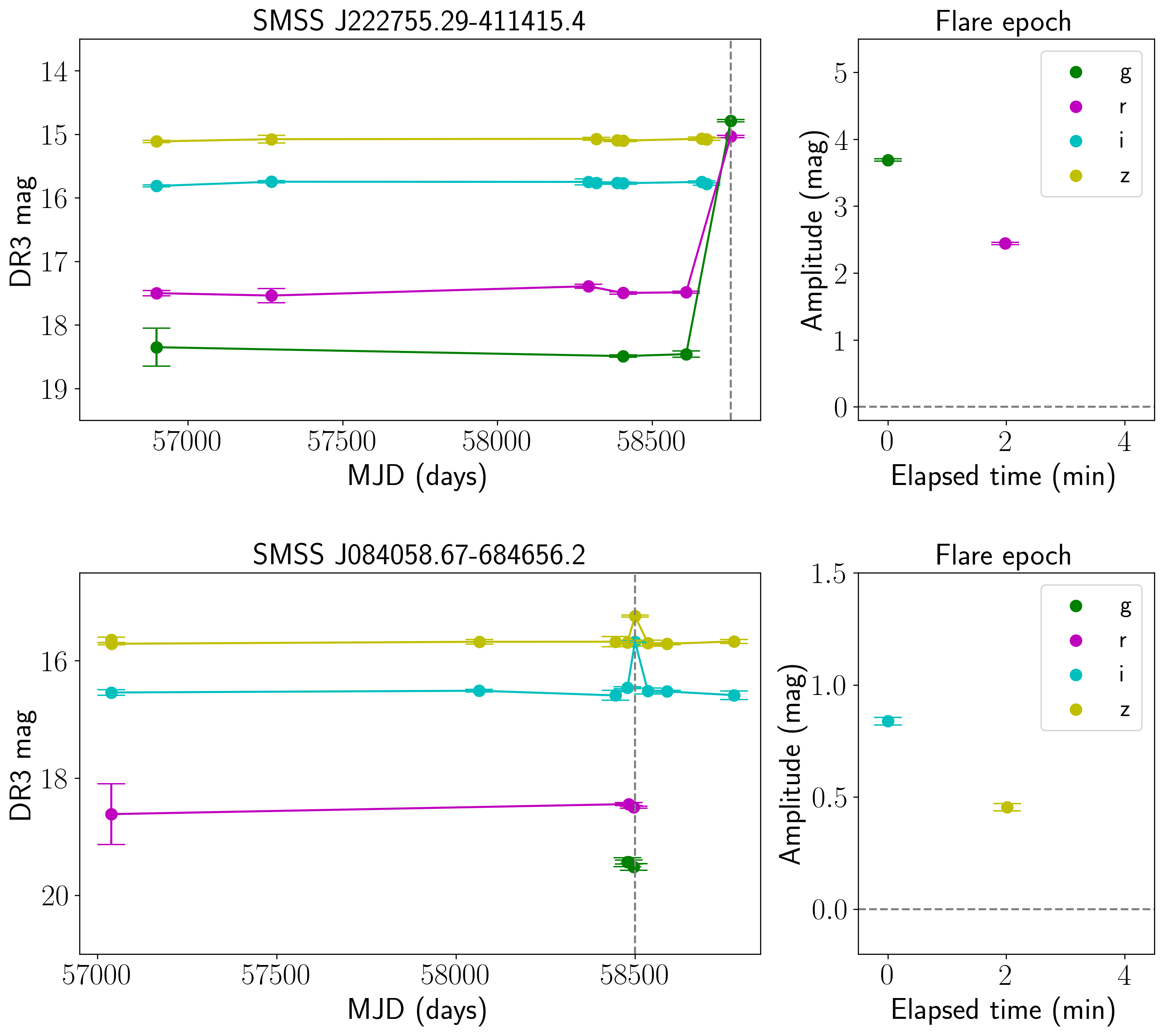}
\caption{Example light curves of flaring M dwarfs in SMSS DR3, labelled with their SMSS names. From top to bottom, the flares are detected by (i) a Shallow Survey filter sequence (\texttt{SMSS J091932.65-133044.5}), (ii) a Main Survey filter sequence (\texttt{SMSS J021804.45-020939.1}), (iii) a \(gr\) filter pair (\texttt{SMSS J222755.29-411415.4}), and (iv) an \(iz\) filter pair (\texttt{SMSS J084058.67-684656.2}). {\refbf For clarity, we increase the size of the error bars by a factor of five for all data points.}}
\label{fig:Examples of DR3 flare lightcurves}
\end{figure}

We use the third data release of the SkyMapper Southern Survey (SMSS DR3) to extract a new, extended sample of M dwarf flares\footnote{DR3 is currently accessible only to Australia-based researchers and their collaborators.}. 
DR3 includes observations from March 2014 to October 2019, more than doubling the deep coverage over the previous data release (DR2: \citealt{Onken2019PASA...36...33O}). With 5-$\sigma$ depth for point sources $\sim$19--21 mag depending on the filter, these deeper images from the Main Survey allow us to cover a larger volume with more distant M dwarfs than our previous work (see Figure 2 of \citealt{Chang2020MNRAS.491...39C}). DR3 also contains 50\% more observations from the Shallow Survey component \citep{Wolf2018PASA...35...10W}. The previous analysis by \citet{Chang2020MNRAS.491...39C} was based on 254 flares found purely in the Shallow Survey data of SMSS DR1.1. The improved data set of DR3 increases the flare numbers including lower-amplitude flares detected in deeper blue-filter \(uv\) observations.

DR3 provides the \texttt{dr3.photometry} table where we construct light curves of M dwarfs in different photometric bands. Due to survey strategy and observational conditions, typical SkyMapper cadences are not uniform in either survey component, and resulting multi-colour light curves have a sparse and uneven data sampling.  For over 600 million unique astrophysical objects, the median number of observations (\texttt{ngood}) with a semi interquartile range is 18$\pm$4 measurements across all filters. If observations with sparse sampling detected a bright state in just a single image, it would not be considered reliable proof of a flare; however, if the brightening was seen in multiple images, it would be convincing.

The SkyMapper survey strategy contains indeed multiple exposures during every visit to a given survey field, and thus provides robust evidence for flare activity. The Shallow Survey uses a filter sequence of the form [\(u\)-\(v\)-\(g\)-\(r\)-\(i\)-\(z\)] with exposure times of 40, 20, 5, 5, 10 and 20 seconds, respectively. In contrast, the Main Survey has a colour sequence of the form [\(u\)-\(v\)-\(g\)-\(r\)-\(u\)-\(v\)-\(i\)-\(z\)-\(u\)-\(v\)] with a uniform exposure time of 100 seconds; this sequence includes three \(u\)--\(v\) pairs taken within a 20 minute-visit to each patch of the sky. The Main Survey also collects additional exposure pairs in \(g\)--\(r\) and \(i\)--\(z\) on different nights. Throughout this paper, we use a general expression for all four different filter sequences referred to as an observation block. Fig.~\ref{fig:Examples of DR3 flare lightcurves} shows example light curves of four M dwarfs that flare in the DR3 data set, showing the benefit of near simultaneous multi-colour observations even in the redder filters.

\begin{figure}
\includegraphics[width=\linewidth]{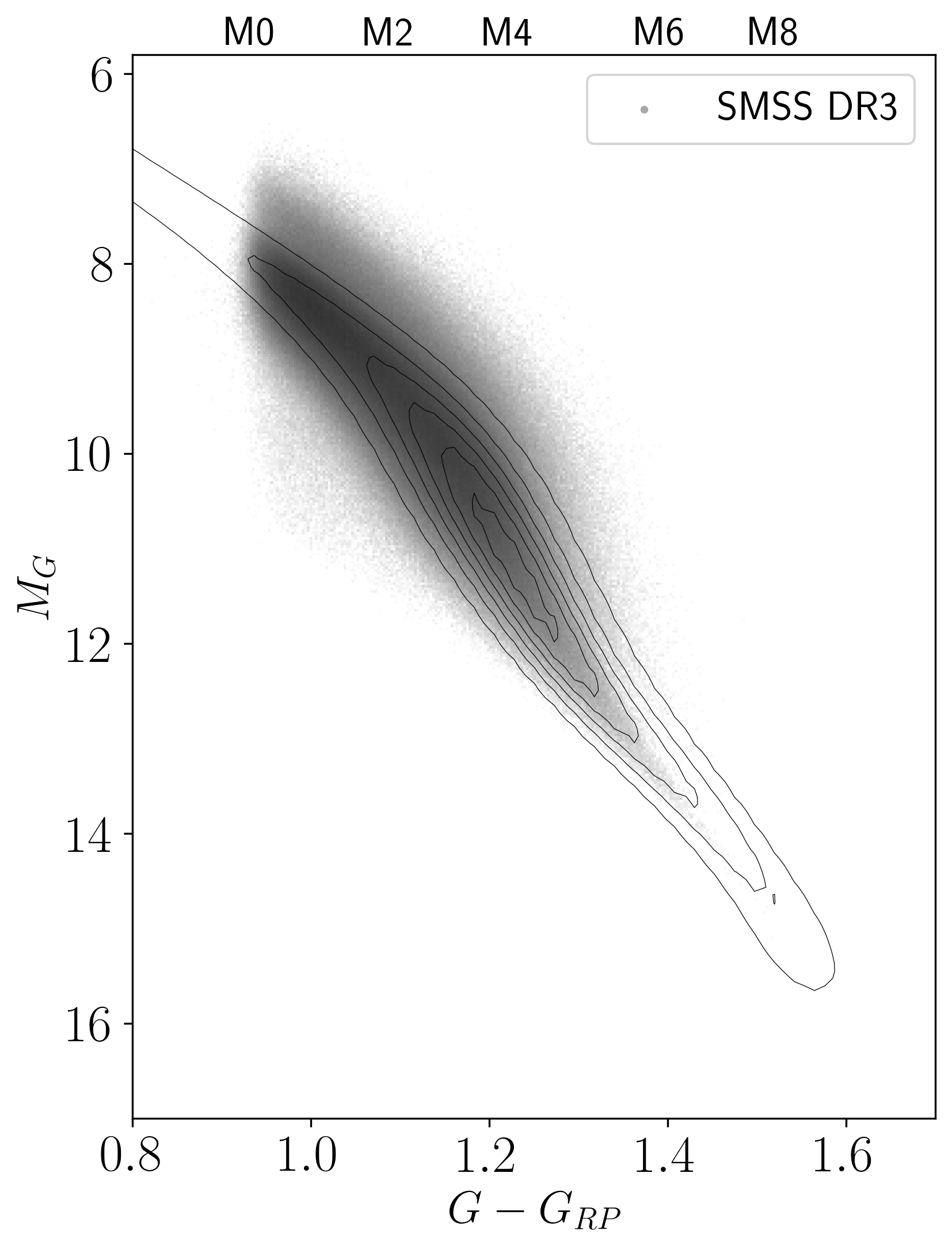}
\includegraphics[width=\linewidth]{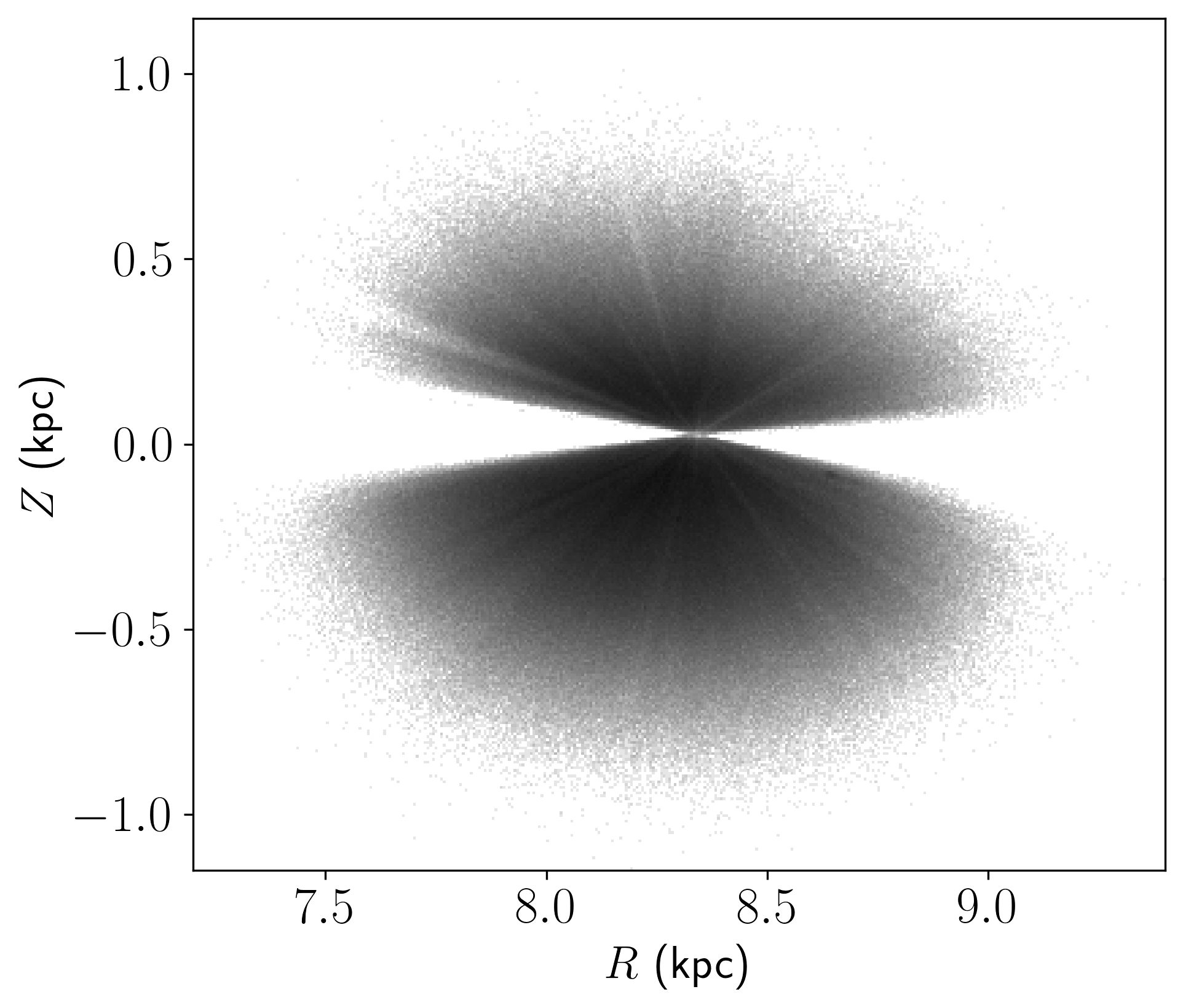}
\caption{Top: SMSS DR3 M dwarf sample in colour-absolute magnitude diagram with Gaia EDR3 photometry. The contour indicates the locus of main sequence stars in the Gaia Catalogue of Nearby Stars. Bottom: Sample distribution in the Galactic cylindrical coordinates (R, Z).}
\label{fig:DR3 M dwarf sample}
\end{figure}

\subsection{Extended M dwarf sample in DR3}
\label{subsec:DR3 M dwarfs}
Following the procedure described in \citet{Chang2020MNRAS.491...39C}, we select an extended sample of M dwarfs in SMSS DR3. First, we apply previous ($r-i$ vs. $i-z$) colour criteria (\texttt{0.75 < $r - i$ < 3.5} and \texttt{0.25 < $i - z$ < 1.7}) and quality cuts (\texttt{FLAGS<4}, \texttt{prox>5}, \texttt{nch\textunderscore max=1}, and \texttt{ebmv\textunderscore sfd<0.2}) to extract an initial sample of objects from the DR3 \texttt{master} table, which results in a $\sim$ 9 times larger sample size than our previous initial sample. One difference is that we make a hard magnitude cut at $r<$~19.5, which is 1.5 mag deeper than our previous work. Next, we remove sources projected near or on top of galaxies and quasars in the spectroscopic catalogues of the 2 Degree Field Galaxy Redshift Survey \citep[\texttt{ext.spec\textunderscore 2dfgrs};][]{2001MNRAS.328.1039C}, the 2dF Gravitational Lens Survey \citep[\texttt{ext.spec\textunderscore 2dflens};][]{2016MNRAS.462.4240B}, the Million Quasar catalog, v.6.4b \citep[\texttt{ext.milliquas\textunderscore v6p4b};][]{2015PASA...32...10F}, the 2dF and 6dF QSO Redshift Surveys \citep[\texttt{ext.spec\textunderscore 2qz6qz};][]{2004MNRAS.349.1397C}, the 6 Degree Field Galaxy Survey \citep[\texttt{ext.spec\textunderscore 6dfgs;}][]{2009MNRAS.399..683J}, the Galaxy Mass and Assembly Survey DR3 \citep[\texttt{ext.spec\textunderscore gama\textunderscore dr3};][]{2018MNRAS.474.3875B}, which we have all pre-matched to the SMSS \texttt{master} table with a 15~arcsec search radius. {\refbf This procedure removes $\sim$0.02\% of sources in the initial SMSS sample of 13.6 million M dwarf candidates.}

Then, we use high-precision astrometric and photometric measurements from Gaia's early data release 3 \citep[EDR3;][]{2016A&A...595A...1G, 2021A&A...649A...1G} to identify likely main sequence stars. After correcting for proper motion, we crossmatch SkyMapper DR3 and Gaia EDR3 with a match radius of 1 arcsec and keep only objects with parallaxes of at least 10$\sigma$ significance from the EDR3. 
Finally, we apply several quality cuts recommended by the Gaia Collaboration to obtain a clean sample with fewer spurious parallax and problematic photometric values. As explained in \citet{2021A&A...649A...3R} and \citet{2021A&A...649A...4L}, we correct the $G$-band magnitude and parallax of a given Gaia EDR3 source before filtering out\footnote{\url{https://www.cosmos.esa.int/web/gaia/edr3-code}}. The corrections for $G$-band magnitude are available for nearly all selected sources but the inclusion of the correction term shows a negligible improvement in overall ($\left| \Delta G \right| \leq$  0.025 mag). The parallax zero-point offset is corrected based on ecliptic latitude, $G$-band magnitude, and colour information, and it results in reducing the inferred distance for most sample stars. The maximum difference is up to --200 pc for stars at distances of about 2 kpc. Note that in this work we use a Bayesian distance with exponentially decreasing prior with scale length of \(L\) = 1 kpc following \citet{2018A&A...616A...9L}. 
The selection criteria are summarised as follows:

\begin{itemize}
    \item \texttt{parallax/parallax\textunderscore error(=$\varpi/\varpi_{e}$)>10}: We adopt a 10\% relative precision criterion which corresponds to an uncertainty on $M_{G}$ smaller than 0.22 mag.\\
    
    \item \texttt{|C$^{*}$|<0.2}: We use the corrected BP and RP flux excess factor, C$^{*}$, suggested in \citet{2021A&A...649A...3R}, which is zero for normal sources with a low blend probability or no variability.\\
    
    \item \texttt{ruwe<1.4}: We use the renormalised unit weight error (RUWE) as recommended in \citet{2021A&A...649A...4L} to exclude sources with poor astrometric solutions or remove potential non-single objects.\\
    
    \item \texttt{ipd\textunderscore frac\textunderscore multi\textunderscore peak$\leq$2}: This new EDR3 parameter is useful for identifying source that are not isolated. We find that most of the sources with a high value of \texttt{ipd\textunderscore frac\textunderscore multi\textunderscore peak} lie on the upper right side of the main sequence at a given colour, where we expect potential binary candidates to exist. \\
         
    \item \texttt{phot\textunderscore bp\textunderscore mean\textunderscore mag-phot\textunderscore rp\textunderscore mean\textunderscore mag>1.8}: We use a simple colour cut $G_\mathrm{BP}-G_\mathrm{RP}>$1.8 to select M dwarfs initially. The main effect of this cut is to remove a small fraction of M0 type stars when we check against $G-G_\mathrm{RP}$ colour later, which has the tightest relation to spectral type for M dwarfs \citep{2019AJ....157..231K}.\\
    
    \item \texttt{M$_{G}$<1.95+(G$_\mathrm{BP}$-G$_\mathrm{RP}$)+1.5}: We remove red giants by placing a diagonal line parallel to main sequence locus on the Gaia's colour-absolute magnitude diagram (CaMD). {\refbf Before applying this selection cut, the M giant contamination rate is about 0.45\% for 3.6 million M dwarf sample that passed above Gaia quality cuts.} \\
    
    \item \texttt{M$_{G}$>1.95+(G$_\mathrm{BP}$-G$_\mathrm{RP}$)+7.0}: This additional line is for excluding a few metal-poor subdwarf candidates (i.e., sdM and esdM types) below the main sequence{\refbf, which further removes about 0.04\% of the objects.}
\end{itemize}

As a result, we obtain a flux-limited sample of 3,616,738 objects with 32,680,080 observation blocks, which are about two and seven times larger than previous work, respectively. Fig.~\ref{fig:DR3 M dwarf sample} shows the comparison between our clean sample of M dwarfs and the Gaia Catalogue of Nearby Stars within 100~pc (\citealt[GCNS;][]{2021A&A...649A...6G}), confirming a good agreement with our selection in the overlapping regime across the lower CaMD. As indicated by broad main sequence locus, our sample should have M dwarfs with a wide range of ages and/or metallicity in the Galactic disc. Since our final sample has a median and median absolute deviation of E(B-V) of 0.057$\pm$0.0303, reddening has no significant effect on $G-G_\mathrm{RP}$ colour. {\refbf Unresolved binaries with different mass ratio can add statistical uncertainty to flare rate estimation due to spectral mis-classification. However, it turns out that their contamination effects are not severe enough to significantly affect our result (see Section 3).} As shown in the bottom panel of Fig.~\ref{fig:DR3 M dwarf sample}, our final sample is restricted to stars around $R=8.34\pm1$~kpc and out to $|Z|\sim1$~kpc away from the Galactic mid-plane, with a maximum distance of $D< 1.5$~kpc. M dwarfs in the analysed volume are well explained by two exponential discs (the thin and thick disc), where we expect populations with an average age of 4--5 Gyr (e.g., \citealt{2008ApJ...673..864J, Casali2019A&A...629A..62C, 2019MNRAS.486.1167B}).

\begin{table}
\caption{Coefficients of Null Distributions and Threshold values for flare candidate detection.}
\centering
\begin{tabular}{cccccc}
\hline \noalign{\smallskip}  
 &   &  &  & \multicolumn{2}{c}{$\Phi$ Threshold at $\alpha\simeq0.05$} \\
\noalign{\smallskip} \cline{5-6} \noalign{\smallskip} 
Survey element & $\Phi_{ij}$ &  $\mu_\mathrm{DR3}$ & $\sigma_\mathrm{DR3}$ & DR1.1 & DR3 \\
\noalign{\smallskip} \hline \noalign{\smallskip}
        & $uv$  & 0.916 & 0.516  & 540 & 260 \\
        & $vg$  & 0.876 & 0.486  & 420 & 220 \\
Shallow & $gr$  & 0.891 & 0.488  & \textbf{880} & \textbf{620} \\
        & $ri$  & 0.886 & 0.485  & 755 & 930 \\
        & $iz$  & 0.893 & 0.483  & 765 & 1360 \\
\noalign{\smallskip} \hline \noalign{\smallskip}
        & $uv$  & 0.902 & 0.528  &  -  & 280 \\
        & $vg$  & 0.698 & 0.534  &  -  & 230 \\
Main    & $gr$  & 0.639 & 0.523  &  -  & \textbf{410} \\
(default) & $ru$  & 0.758 & 0.571&  -  & 360 \\
        & $vi$  & 0.705 & 0.551  &  -  & 360 \\
        & $iz$  & 0.692 & 0.509  &  -  & 980 \\
        & $zu$  & 0.737 & 0.562  &  -  & 370 \\
\noalign{\smallskip} \hline \noalign{\smallskip}
Main    & $gr$  & 0.626 & 0.521 & - & \textbf{390} \\
(extra) & $iz$  & 0.713 & 0.515 & - & 1180 \\
\noalign{\smallskip} \hline
\end{tabular}
\label{tab:tab1}
\end{table}

\section{Method}
\label{sec:method}
We use near-simultaneous broadband data to identify flare candidate events in the same way as done for SMSS DR1.1 (see \citealt{Chang2020MNRAS.491...39C}). The flare variability index $\Phi$ for any consecutive filters (\(i\), \(j\)) in each observation block \(n\) is defined as:
\begin{align}
\Phi_{ij,n} =  [\frac{m_{i,n} - \overline{m}_{i}}{\sigma_{i,n}}] [\frac{m_{j,n} - \overline{m}_{j}}{\sigma_{j,n}}] ~,
\end{align} 
where $m_{i,j}$ are single measurement points (\texttt{mag\textunderscore psf}) for each filter at epoch $n$, $\sigma_{i,j}$ are their errors (\texttt{e\textunderscore mag\textunderscore psf}), and $\overline{m}_{i,j}$ are the quiescent magnitudes of those filters. In case of pure noise, this index is close to a normal distribution with a mean of $\Phi_{ij} \approx 0$. Given there are more measurements in DR3 than before, we can estimate the quiescent magnitude more reliably than before, using our leave-one-out strategy.

\begin{figure}
\includegraphics[width=\linewidth]{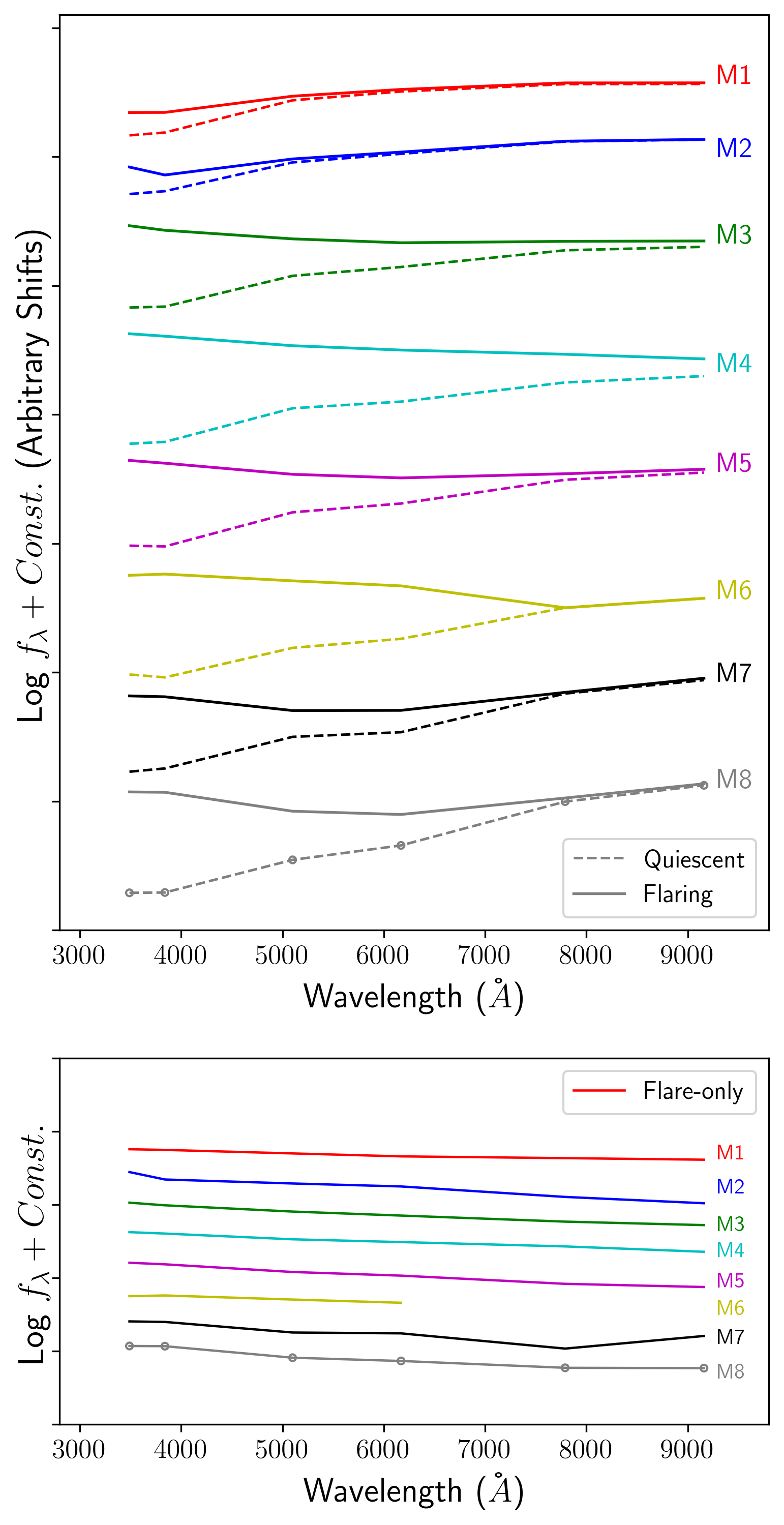}
\caption{Top: Flaring (solid lines) and quiescent (dashed lines) SEDs of M dwarfs with large continuum enhancement in the \(uv\) bands. Solid circles indicate the location of the central filter wavelength of \(u, v, g, r, i\), and \(z\) bands. Bottom: Flare-only SEDs for the above cases.}
\label{fig:flare SED}
\end{figure}

We initially select flare candidates that meet the following conditions: \texttt{$m_{i,n} - \overline{m}_{i} < 0$ AND $m_{j,n} - \overline{m}_{j} < 0$ AND} \texttt{$\Phi_{ij}$ > 0}. To reduce the number of false positives caused by correlated random noise in our data, we compare the flare candidate distribution (\texttt{$\Phi_{ij}$ > 0}) with a null distribution (\texttt{$\Phi_{ij}$ < 0}) adopting the method of \citet{2001AJ....122.3492M}. The null distribution for each filter pair is well described by the probability density function of a normalised Gaussian whose mean $\mu$ and standard deviation $\sigma$ are given in Table \ref{tab:tab1}. The last column in the table gives a threshold value corresponding to a contamination fraction of only 5\% ($\alpha=0.05$) for individual filter pairs in different survey modes. The reason we specify the survey mode for the same filter pairs is that the noise in SkyMapper light curve data depends on observing conditions and exposure times. We already know that the shape of the noise distribution has a significant effect on our correlation-based selection method to flare identification, especially for ones with smaller amplitudes. We demonstrate this effect by comparing the threshold values of \(gr\)-filter pairs obtained in four different cases (marked in bold in Table \ref{tab:tab1}). Since photometric measurement data in DR3 have a sharper noise distribution than DR1.1, it tends to lower the value of a noise threshold cut. The gain in imaging depth of SkyMapper Main Survey pushes this limit further. Moreover, as we restricted additional \(gr\)-filter pairs under dark sky conditions mostly \citep{Onken2019PASA...36...33O}, it provides extra advantage of reducing the random photometric errors. 
To improve the purity of the initially selected sample, it is necessary to remove candidates that have a low signal-to-noise ratio (SNR) and/or small-changes in amplitude. The SNR of each candidate signal is calculated as $\Delta m / \sigma_{LC}$, where $\Delta m (=|m_{n} - \overline{m}|)$ is the flare amplitude measured relative to the quiescent magnitude and $\sigma_{LC}$ is the light curve root-mean-square scatter. We adopt the same SNR threshold of 3.0 to all filter pairs as a conservative approach. To collect a reliable sample that follows the pattern of a typical flare variation (see Section \ref{subsec:DR3 lightcurves}), we use only candidates with amplitude $>$ 0.2 mag and eyeball them all for a quality check. As a result, we find 1,379 candidate epochs among 42,079,396 observed epochs in any filter combinations, which contains 933 unique flare candidates (see the last column). As shown in the Table \ref{tab:tab2}, it is clear that \(uv\)-filter pairs are very helpful and sensitive for a flare search but its ability is mainly limited by number of observed epochs due to faintness of the spectral energy distribution of M dwarf in the blue wavelengths. This limitation is inherent to any blue-red filter pairs such as \(ru\), \(vi\) and \(zu\) due to the same reason. While the reddest filter pair \(iz\) is the least sensitive for the flare search, the mid-range filter pair \(gr\) provides the main contribution to our sample because of its observed depth and hence sensitivity for detecting relatively low-amplitude flares. Fig.~\ref{fig:flare SED} shows the contrast of the flare spectral energy distribution (SED) against the quiescent (non-flare) SED for early to late M dwarfs with high-amplitude flares ($\Delta m >1$~mag). M dwarfs with spectral subtype M3 or later have a good visibility in the \(uvg\) bands. The flare-only SEDs around peak emission are intrinsically similar in nature across spectral types from M1 to M8, and resemble a general shape of a blackbody with temperatures of $T\sim$ 9\,000--10\,000 K.

\begin{table*}
\caption{Summary of flare candidates from SkyMapper DR3. The similar analysis of \citet{Chang2020MNRAS.491...39C} used 254 flares from SkyMapper DR1.1.}
\centering
\begin{tabular}{ccccccc}
\hline \noalign{\smallskip}  
&  &   & \multicolumn{4}{c}{Number of flare epochs} \\
\noalign{\smallskip} \cline{4-7} \noalign{\smallskip} 
Survey Component & $\Phi_{ij}$ &  Number of observed epochs & $\alpha \simeq 0.05$ & SNR $\geq 3$ & $\Delta m > 0.2$ \& Eyeball & Unique Candidates\\
\noalign{\smallskip} \hline \noalign{\smallskip}
        & $uv$  &     27,955 &   198 &   133 & 103 & 103\\
        & $vg$  &     30,544 &   183 &   118 & 101 &  21  \\ 
Shallow & $gr$  &  2,712,767 &   572 &   336 & 224 & 170  \\ 
        & $ri$  &  6,539,637 &    278 &  165 & 119 &  35   \\ 
        & $iz$  & 13,157,599 &   143 &   72 &  50 &  11   \\
\noalign{\smallskip} \hline \noalign{\smallskip}
        & $uv_\mathrm{first}$  &  40,041 & 245 & 159 & 65 & 65 \\
        & $vg$ & 63,424 & 187 & 158 & 71 & 37 \\
        & $gr$ & 2,606,385 & 1007 & 585 & 153 & 114\\
        & $ru$ & 20,901 & 56 & 23 & 19 & 4 \\        
Main    & $uv_\mathrm{middle}$  & 27,340 & 168 & 128 & 38 & 12 \\
(default) & $vi$  & 41,803 & 63 & 57 & 39 & 8 \\
        & $iz$  & 2,690,922 & 52 & 19 & 15 & 2 \\
        & $zu$  & 17,436 & 29 & 8 & 7 & 2\\        
        & $uv_\mathrm{last}$  & 18,326 & 123 & 95 & 31 & 5\\        
\noalign{\smallskip} \hline \noalign{\smallskip}
Main   & $gr$   & 4,657,456 & 1,882 & 1,045 & 297 & 297 \\
(extra) & $iz$  & 9,426,860 & 191 & 115 & 47 & 47 \\
\noalign{\smallskip} \hline \noalign{\smallskip}
Total       &        & 42,079,396  & 5,377 & 3,216 & 1,379  & 933\\
\noalign{\smallskip} \hline \noalign{\smallskip}
After cleaning      &        &             &       &      &        & 906 \\
\noalign{\smallskip} \hline
\end{tabular}
\label{tab:tab2}
\end{table*}

Finally, we check any contamination by known solar system objects and variable stars. For the former case, we use the SkyBoT cone-search service \citep{2006ASPC..351..367B} to remove candidate epochs where the observed changes in the apparent magnitude caused by known asteroids. Among 933 unique flaring epochs, we find that 27 candidates turn out to be false positives due to the effect of source blending. In the case of the latter, 39 of our flaring stars are recognised as known variable stars listed in the AAVSO International Variable Star Index \citep[v.2019-09-30;][]{2006SASS...25...47W,2017yCat....102027W}, which is cross-matched to DR3 in table \texttt{ext.vsx\textunderscore 20190930}. Except for a few variables of unspecified type, most of them are classified as one of three types: 20 spotted stars whose rotational variability is closely related to stellar activity, five classical flare stars such as UV Ceti, and active six M dwarfs in the binary systems. For this work we decide to include all these classes of activity-related variables. It thus provides us the final sample of 906 unique flare events.

{\refbf To understand the effect of binary contamination, we use the non-single star tables newly added in the Gaia DR3 catalogue \citep{Halbwachs2022arXiv220605726H}. We find 17 stars with an orbital two-body solution (13 Eclipsing Binaries, 1 Single Lined Spectroscopic binary (SB1), and 3 Double Lined Spectroscopic binaries) and one SB1 candidate compatible with a first-degree trend. This implies a sample contamination rate of 0.0005\% as a lower limit on the unresolved binary fraction. We also identify 429 unresolved WD+dM candidates (White-Dwarf+M dwarf pair) by following the procedure outlined by \citet{Smolcic2004ApJ...615L.141S}, requiring a colour cut of $u - g < 2.35$. Again the sample contamination rate is low enough (only 0.01\%). None of these non-single stars have flaring epochs. Furthermore, we consider those of resolved binary stars within $\sim$1 kpc of the Sun based on compilation by \citet{El-Badry2021MNRAS.506.2269E}. By crossmatching our DR3 M dwarfs to this catalogue, we identify 64,567 (59,472) high-confidence binary pairs with 90\% (99\%) probability of being bound, including 48,586 (43,740) main-sequence -- main-sequence (MSMS) binaries, 2,452 (2,274) WDMS binaries, and 13,529 (13,458) MS?? (MS + one with no colour) binaries. Among them, only 30 (29) M dwarfs in such binaries shows flare activity that is about 3.3\% (3.2\%) of the result in the final flare sample. This also allows us to define flare sample with a very low contamination rate of 0.1\% by WDMS pairs. The range of projected separation of these binaries is between 46 AU and 127,442 AU, so that it is reasonable to assume that the properties of the M dwarf in these wide binaries should be similar to those of single M dwarfs (i.e., unaffected by the presence of companion).
}

\begin{figure}
\includegraphics[width=\linewidth]{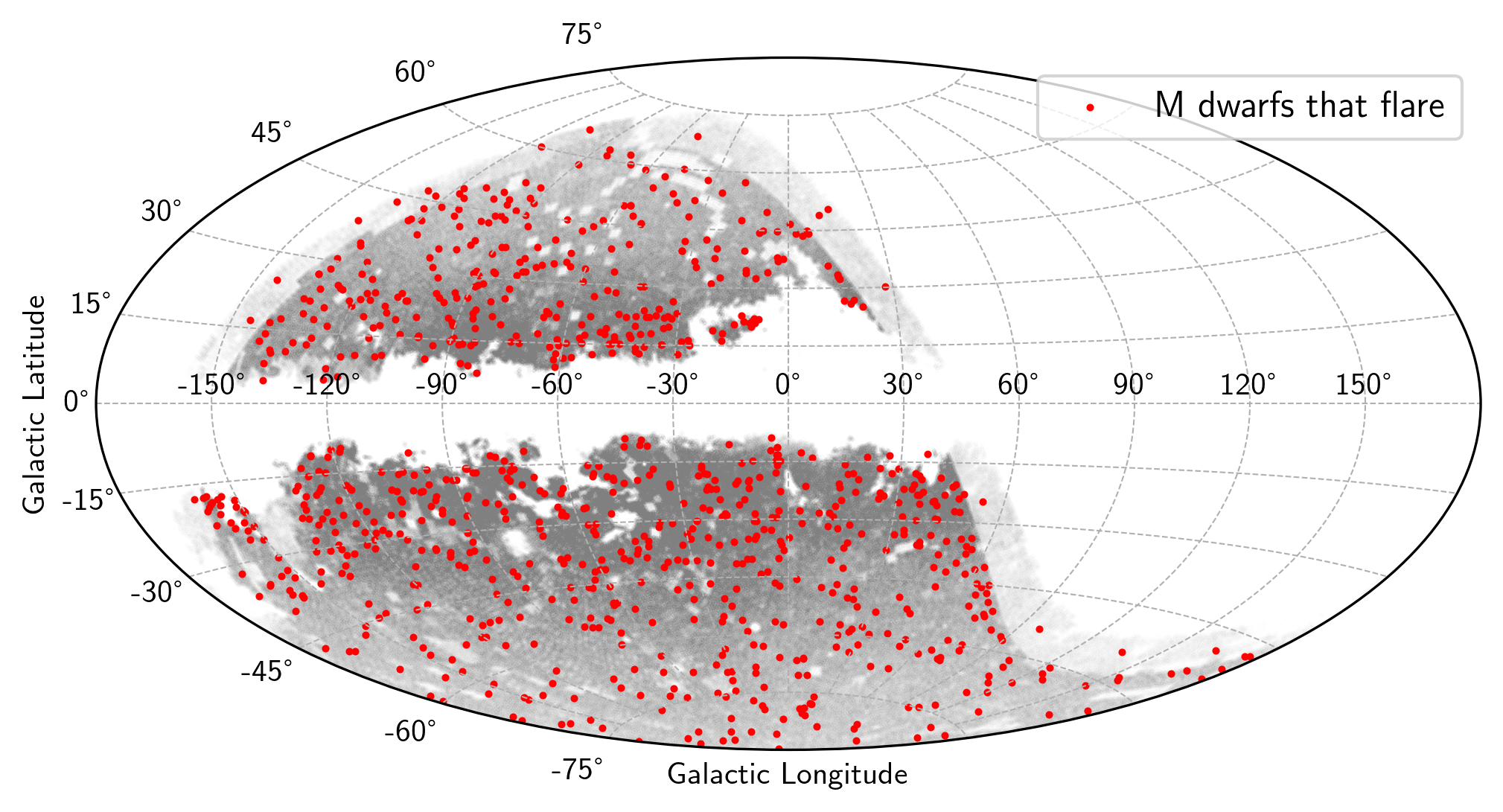}
\caption{Sky coverage of selected M dwarf sample (gray) in SMSS DR3 and those that flare (red). In this analysis, we reject higher-reddening regions and flagged SkyMapper fields (white).}
\label{fig:sky coverage}
\end{figure}

\section{Results and Discussion}
\label{sec:results and discussion}
With the benefit of increased data volume and deep sky coverage in DR3, the number of M dwarf flares is increased by factor of 3.5 compared to our previous work (See Table 2 of  \citealt{Chang2020MNRAS.491...39C}). Both the observed M dwarfs and a subset of stars that flare are distributed over the sky in an unbiased way  (Fig.~\ref{fig:sky coverage}). We still avoid high-reddening regions with E(B-V) $\geq 0.2$, especially for Galactic latitude $|b|<10$, where the survey did not address detailed issues of source de-blending, extinction correction, and photometry. The forthcoming SMSS data release (DR4) is planned to overcome some of these issues and may enable finding flares also in the zone of young stars and from old binary systems in the Galactic bulge.

In this section, we show how we extend our previous analysis of flare stars to include additional survey data with near-simultaneous measurements in four different modes (see Section \ref{subsec:DR3 lightcurves}). We first investigate possible selection bias due to filter-dependent variability of M dwarf flares (Section \ref{sec:survey bias}). Because of the same flare amplitude cut, flares with low energy release are mostly detectable in our bluest \(uv\) filter pairs, while in the redder ones the number census of the selected flare sample is highly biased toward very high-energy flares or flares seen near peak phase in the light curves. Thanks to the increased sample size of M dwarfs and flares, we can now greatly improve the resolution of flaring fraction in bins of spectral type (Section \ref{sec:spatially dependent flaring fraction}), Galactic height above or below the plane, $Z$, (Section \ref{sec:Z-dependent flaring fraction}), and radial and vertical (R, Z) position in the Galaxy (Section \ref{sec:(R, Z)-dependent flaring fraction}). To derive robust statistics on the flare rate, the basic assumption is that all stars in the same bin share similarities in their physical properties but also the level of magnetic activity. 

For the sake of clarity, throughout this paper we use ($G - G_\mathrm{RP}$) colour as a good indicator of spectral type. We initially adopt the mean colours of M dwarfs calculated by \citet{2019AJ....157..231K}, which was based on colour information from the Gaia DR2 catalogue. However, we find the presence of small colour differences in our Gaia EDR3 sample when we plot the colour versus absolute magnitude diagram using Gaia bands. With a recent compilation of spectroscopic M dwarf catalogues by \citet{2019AJ....157..231K}, we re-calculate Gaia EDR3 mean colours and absolute magnitudes for M dwarfs as a function of spectral subtype (M0 to M9). Table \ref{tab:tab3} summarises the mean ($G - G_\mathrm{RP}$) colour $\mu$, its standard deviation $\sigma$, and colour offset between the two catalogues. Although the offset is small, it systematically increases from early- to late-type M dwarfs. For the $G$-band, the mean magnitude and its standard deviation per spectral type is fully consistent with the recent Gaia DR2 result (see Table 4 of \citealt{2019AJ....157..231K})\footnote{For the latest filter throughput and CCD response data, see \url{https://www.cosmos.esa.int/web/gaia/edr3-passbands}}. 

\begin{table}
\caption{Mean Gaia $G - G_\mathrm{RP}$ colour for M dwarfs in DR2 and EDR3.}
\centering
\begin{tabular}{cccccccc}
\hline \noalign{\smallskip}
&  \multicolumn{2}{c}{$\mathrm{DR2}$} & &\multicolumn{2}{c}{$\mathrm{EDR3}$} & & $\mathrm{EDR3 - DR2}$ \\
\noalign{\smallskip} \cline{2-3} \cline{5-6} \cline{8-8} \noalign{\smallskip} 
Type & $\mu$ &  $\sigma$ &  & $\mu$ &  $\sigma$ & & $\Delta \mu$\\
\noalign{\smallskip} \hline \noalign{\smallskip}
M0   & 0.93 & 0.04 &  & 0.92  & 0.04 & & $-$0.01 \\
M1   & 1.01 & 0.04 &  & 1.00  & 0.04 & & $-$0.01\\
M2   & 1.09 & 0.03 &  & 1.08  & 0.03 & & $-$0.01\\
M3   & 1.16 & 0.03 &  & 1.14  & 0.03 & & $-$0.02\\
M4   & 1.23 & 0.04 &  & 1.21  & 0.04 & & $-$0.02\\
M5   & 1.32 & 0.05 &  & 1.30  & 0.05 & & $-$0.02\\
M6   & 1.41 & 0.04 &  & 1.38  & 0.04 & & $-$0.03\\
M7   & 1.47 & 0.05 &  & 1.43  & 0.05 & & $-$0.04\\
M8   & 1.57 & 0.05 &  & 1.51  & 0.04 & & $-$0.06\\
M9   & 1.63 & 0.05 &  & 1.55  & 0.04 & & $-$0.08\\
\noalign{\smallskip} \hline
\end{tabular}
\label{tab:tab3}
\end{table}

\subsection{Investigation of Systematic Effects}
\label{sec:survey bias}

\begin{figure*}
\includegraphics[width=\linewidth]{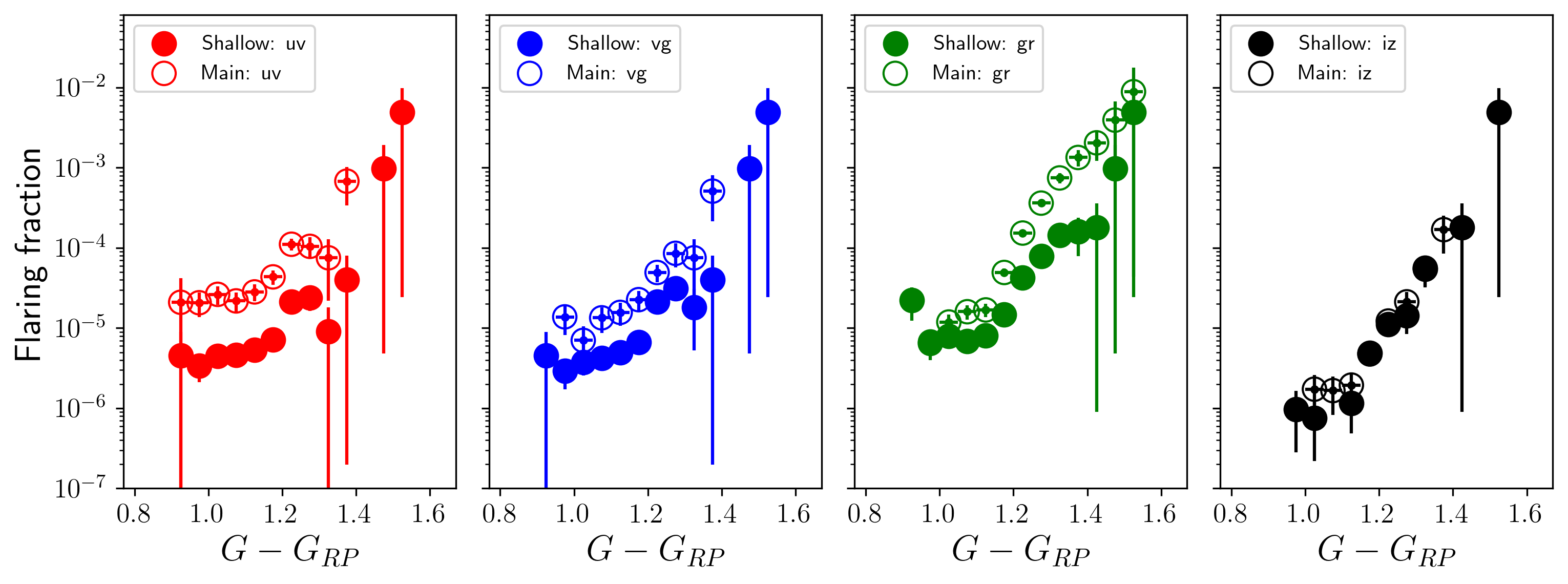}
\caption{Flaring fraction of the shallow survey (solid circles) and main survey (open circles) for different filter pairs. The horizontal error bars are bin width (0.05 mag) and vertical errors bars are Poisson uncertainties.}
\label{fig:survey-dependent flaring fraction}
\end{figure*}

To understand the effect of our selection criteria ($\Delta m > 0.2$ mag), we first check how the resulting distribution of flaring fraction changes with survey parameters (e.g., depth and filter sequence). Here we define the flaring fraction as the ratio between the number of flare epochs (or observation blocks) and the total number of observed epochs as a function of ($G-G_\mathrm{RP}$) in bins of 0.05~mag, which has a factor of 2.5 times better resolution than those in our analysis of DR1.1. The new sample extends our active M dwarf search to later spectral types (M7 to M9), e.g., the closest and reddest M dwarf LP 944-20, also known as \texttt{SMSS J033935.65-352539.5}, has a Simbad spectral classification (M9Ve) with a weak H${\alpha}$ emission \citep{1998MNRAS.296L..42T, 2002A&A...390L..27M}. In Fig.~\ref{fig:survey-dependent flaring fraction}, we compare the results between two survey components with different filter pairs such as \(uv\), \(vg\), \(gr\), and \(iz\). There is a clear trend of increasing flaring fraction towards redder colour common to all of them, giving us almost identical results as using the whole sample discussed below. But the integration times clearly affect the recovery of low-amplitude flares that are mainly limited by the photometric precision of a given survey component. Our Main Survey component with longer exposure can detect more low-luminosity flares than those in the Shallow Survey component with shorter exposures. Since the observed frequency of flares increases with decreasing flare luminosity (e.g., \citealt{Kowalski2009AJ....138..633K, Davenport2012ApJ...748...58D}), it is natural to expect that we have seen a relatively low flaring fraction in the Shallow Survey sample. But this trend is not evident in the \(iz\) filter pairs, which is likely due to the weak contrasting effect of flares in the red-optical band. While in the cases of very high luminosity flares, our blue-band searches sometimes would result in a null signal due to saturation in the SkyMapper images. So, one important gain of our red-optical observations is that we can complement incomplete observations at the brightest end of the flare frequency distribution.

\begin{figure}
\begin{center}
\includegraphics[width=\linewidth]{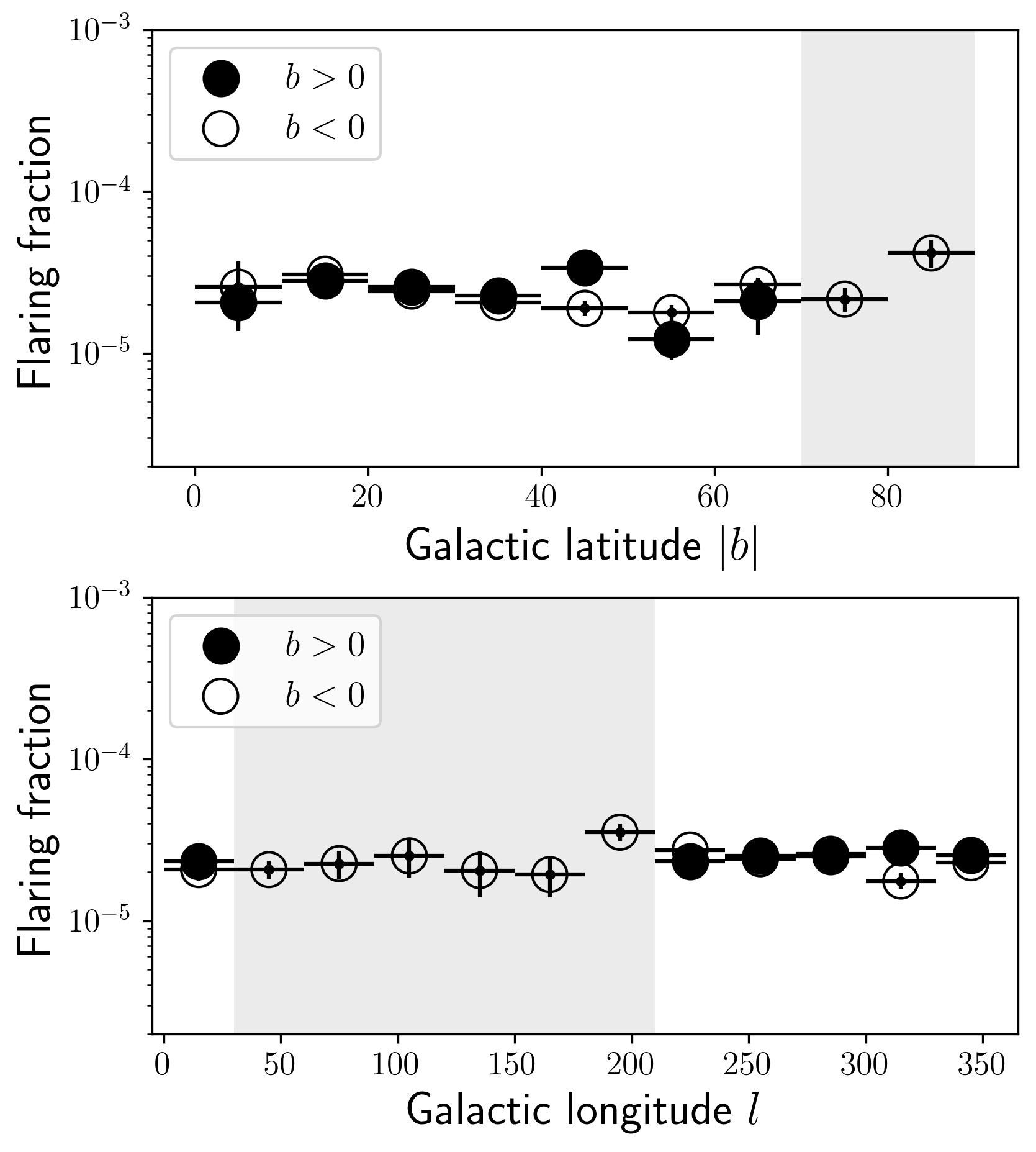}
\caption{M dwarf flaring fraction as a function of Galactic coordinate: we compare the two hemispheres above (solid circles) and below (open circles) the Galactic plane and find no difference between them. Top: dependence on latitude $b$ (bin size 10$\degr$); values at $|b| < 10\degr$ are uncertain due to small number statistics. Bottom: dependence on longitude $l$ (bin size 30$\degr$). The shaded grey areas have no Northern data for comparison (see Fig \ref{fig:sky coverage}). } \label{fig:survey-dependent flaring fraction II}
\end{center}
\end{figure}

Since the flaring fraction may depend on which part of the Milky Way is surveyed, we investigate how it changes with Galactic latitude $b$ and longitude $l$. Note that we did combine flaring epochs detected in the any filter combinations in the sample to increase the SNR. In Fig.~\ref{fig:survey-dependent flaring fraction II}, we find no trend with either coordinate axis, consistent with the results in \citet{Chang2020MNRAS.491...39C}, even close to the low Galactic latitude ($|b| < 10\degr$). {\refbf Our final sample lists 308 flaring blocks over 10,167,978 observation blocks in the northern hemisphere ($b >0$), while in the southern hemisphere ($b <0$), it includes 604 flaring blocks over 22,512,102 observation blocks. The overall observed fraction of flares is basically the same for both hemispheres.} The fraction of M dwarfs seen as flaring with $\Delta m_{uv} > 0.2$ mag in a snapshot of the sky is nearly constant, $\sim 24$ per millions stars, regardless of the line of sight {\refbf (see Table \ref{tab:tab4} for details)}.


\begin{table*}
\caption{Flaring fraction for M dwarfs observed in different ranges of Galactic latitude $b$ for northern and southern hemispheres.}
\centering
\begin{tabular}{cccccc}
\hline \noalign{\smallskip}
Hemisphere & Galactic latitude & Number of stars & All observed blocks & All flaring blocks &  Flaring fraction per million stars \\
\noalign{\smallskip} \hline \noalign{\smallskip}
& $|b| < 10\degr$  & 41,579  & 437,044 &  9  &  $20.6\pm6.7$ \\ 
& $10\degr\leq |b| <20\degr$  & 286,302  & 3,632,274  & 87  &  $28.1\pm2.8$ \\ 
& $20\degr\leq |b| <30\degr$  & 284,271 & 3,689,050 & 79  &  $25.8\pm2.6$   \\ 
& $30\degr\leq |b| <40\degr$  &  225,196 & 2,900,372 & 56  &  $22.8\pm2.8$   \\ 
North & $40\degr\leq |b| <50\degr$  & 153,429 & 1,977,160 &  56  &  $33.9\pm4.1$   \\ 
& $50\degr\leq |b| <60\degr$  & 89,407 & 1,225,405  &  14  &  $12.2\pm3.1$   \\ 
& $60\degr\leq |b| <70\degr$  & 33,896 & 335,315 &  7  &  $20.9\pm7.9$   \\ 
& $70\degr\leq |b| <80\degr$  & 2,869 & 10,742 &  0  & -   \\ 
& $80\degr\leq |b| \degr$  & - &  - &   -  &  -   \\ 
\noalign{\smallskip} \hline \noalign{\smallskip}
& Total & 1,116,949 & 14,207,362 & 308 &  $25.4\pm1.3$  \\ 
\noalign{\smallskip} \hline \noalign{\smallskip}
& $|b| < 10\degr$  & 17,015  & 195,226 & 5 &  $25.6\pm11.5$ \\ 
& $10\degr\leq |b| <20\degr$ &  351,982 & 4,584,525 & 111  &  $30.8\pm2.6$ \\ 
& $20\degr\leq |b| <30\degr$  & 494,271 & 6,559,680 &  129  &  $24.2\pm1.9$   \\ 
& $30\degr\leq |b| <40\degr$  & 461,356 & 5,880,103 &  107  &  $20.6\pm1.9$   \\ 
South & $40\degr\leq |b| <50\degr$  & 395,214 & 4,908,091 &  76  &  $18.9\pm2.0$   \\ 
& $50\degr\leq |b| <60\degr$  & 333,924 & 4,099,957 &  71  &  $17.8\pm2.1$   \\ 
& $60\degr\leq |b| <70\degr$  & 253,186 & 3,204,171  & 55  &  $26.5\pm2.9$   \\ 
& $70\degr\leq |b| <80\degr$  & 144,249 & 1,713,640 &  33 &  $21.6\pm3.5$   \\ 
& $80\degr\leq |b| \degr$  & 48,407 & 624,661 &   17  &  $41.6\pm8.2$   \\ 
\noalign{\smallskip} \hline \noalign{\smallskip}
& Total   & 2,499,604 & 31,770,054  & 604  &  $23.3\pm0.9$  \\ 
\noalign{\smallskip} \hline
\end{tabular}
\label{tab:tab4}
\end{table*}


\subsection{Spatially Dependent Flaring Fraction}
\label{sec:spatially dependent flaring fraction}
To make our results comparable with other studies in the literature, it is important understand how the flare sample is observed. The SMSS DR3 survey volume is much larger than those used in previous photometric (e.g., \citealt[SDSS Stripe 82;][]{Kowalski2009AJ....138..633K}, \citealt[SDSS DR7+2MASS;][]{Davenport2012ApJ...748...58D}, \citealt[Evryscope+TESS;][]{Howard2019ApJ...881....9H}, \citealt[ASAS-SN;][]{Rodriquez2020ApJ...892..144R}) and spectroscopic (e.g., \citealt[SDSS DR7;][]{West2011AJ....141...97W,Pineda2013AJ....146...50P}, \citealt[CARMENES;][]{Jeffers2018A&A...614A..76J}, \citealt[LAMOST DR7;][]{Zhang2021ApJS..253...19Z}) works. The considered area is significantly expanded to include 4\,827 SkyMapper fields covering a total area of over 24\,000 deg$^2$, which is $\sim$1.4 times larger than that used in DR1.1 \citep{Chang2020MNRAS.491...39C}. For example, the main sample used in the SDSS DR7 volume is concentrated on the North Galactic Cap, and does not represent a complete selection of local M dwarfs in the Northern hemisphere. Also, the SDSS Stripe 82 covers a much smaller area, a 2.5 degree wide stripe along the Celestial Equator in the Southern Galactic Cap. High-cadence photometric surveys (e.g., Evryscope and ASAS-SN) with an array of small telescopes are an efficient way of searching for serendipitous flare emission from an all-sky sample of M dwarfs but are limited to magnitudes brighter than $g^{\prime}$=16 or $V$=12--16 mag. 

\begin{figure}
\begin{center}
\includegraphics[width=\linewidth]{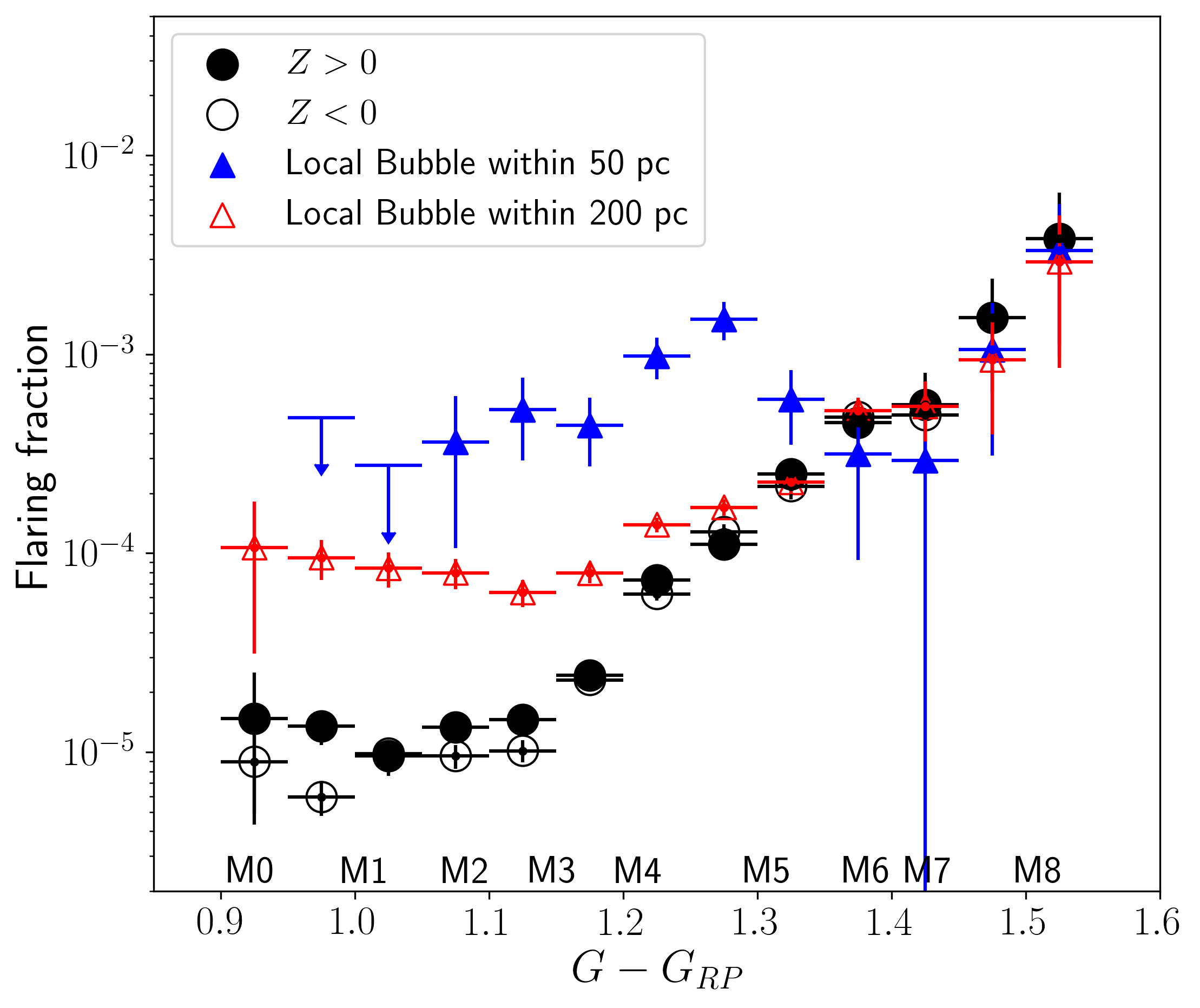}
\caption{Flaring fraction vs. Gaia EDR3 colour for the subsample with the vertical distance above (black solid circles) and below (black open circles) the Galactic plane: horizontal error bars are bin width and vertical errors bars are Poisson uncertainties. For comparison, we plot the volume-limited samples of solar neighbourhood stars out to 50~pc (blue triangle) and 200~pc (red triangle). The spectral types of M dwarfs for M0, M2, M4, M6, and M8 are denoted on the horizontal axis.} 
\label{fig:Flaring fraction vs. Gaia colour}
\end{center}
\end{figure}

To check for additional selection effects, we compare the fraction of M dwarfs that flare for the two subsamples with the vertical distance above ($Z > 0$) or below ($Z < 0$) the Galactic plane at a given colour bin. Following our previous work, we compute the Galactic cylindrical coordinates ($R,\phi,Z$) where the radial distance of the Sun from the Galactic centre is $R_{\odot}=8.34$~kpc and the height of the Sun above the plane is $Z_{\odot}=27$~pc \citep{Gaia2018A&A...616A..11G}. Fig.~\ref{fig:Flaring fraction vs. Gaia colour} shows a typical example of how activity fraction changes with spectral type, i.e., higher flaring fraction increasing spectral type from early to late. Although the total number of observed epochs are a bit different between the subsamples, we see a similar trend and this agrees with the previous results of \citet{Kowalski2009AJ....138..633K}, \citet{West2011AJ....141...97W}, \citet{Jeffers2018A&A...614A..76J} and \citet{Chang2020MNRAS.491...39C} with various activity indicators. The flaring fraction increases steeply from spectral type $\sim$ M4 ($G - G_\mathrm{RP}$ $\sim 1.21$), which is a general trend of field M dwarfs where old stellar population is expected to be dominant. As the flaring fraction varies with stellar age, we consider a reference sample of our solar neighbourhood within the Local Bubble, which includes almost all the stellar nurseries near the Sun \citep{Zucker2022Natur.601..334Z}. The maximum vertical distances $|Z|_\mathrm{max}$ are 74~pc and 218~pc for volume-limited samples with $d<50$~pc and $d<200$~pc, respectively. The stars in these subsamples could also reflect the underlying fraction of flaring activity in a restricted subsample with $|Z|_\mathrm{max}$ cut. The results of the volume-limited samples in Fig.~\ref{fig:Flaring fraction vs. Gaia colour} show that (i) stars close to the Galactic plane show relatively little variation in activity level across spectral type, and (ii) the age effect is particularly clear for the early-type M dwarfs (M0--M3) due to their finite activity lifetimes (e.g., \citealt{West2008AJ....135..785W, Kiman2021AJ....161..277K}). 

\begin{figure*}
\begin{center}
\includegraphics[width=0.98\linewidth]{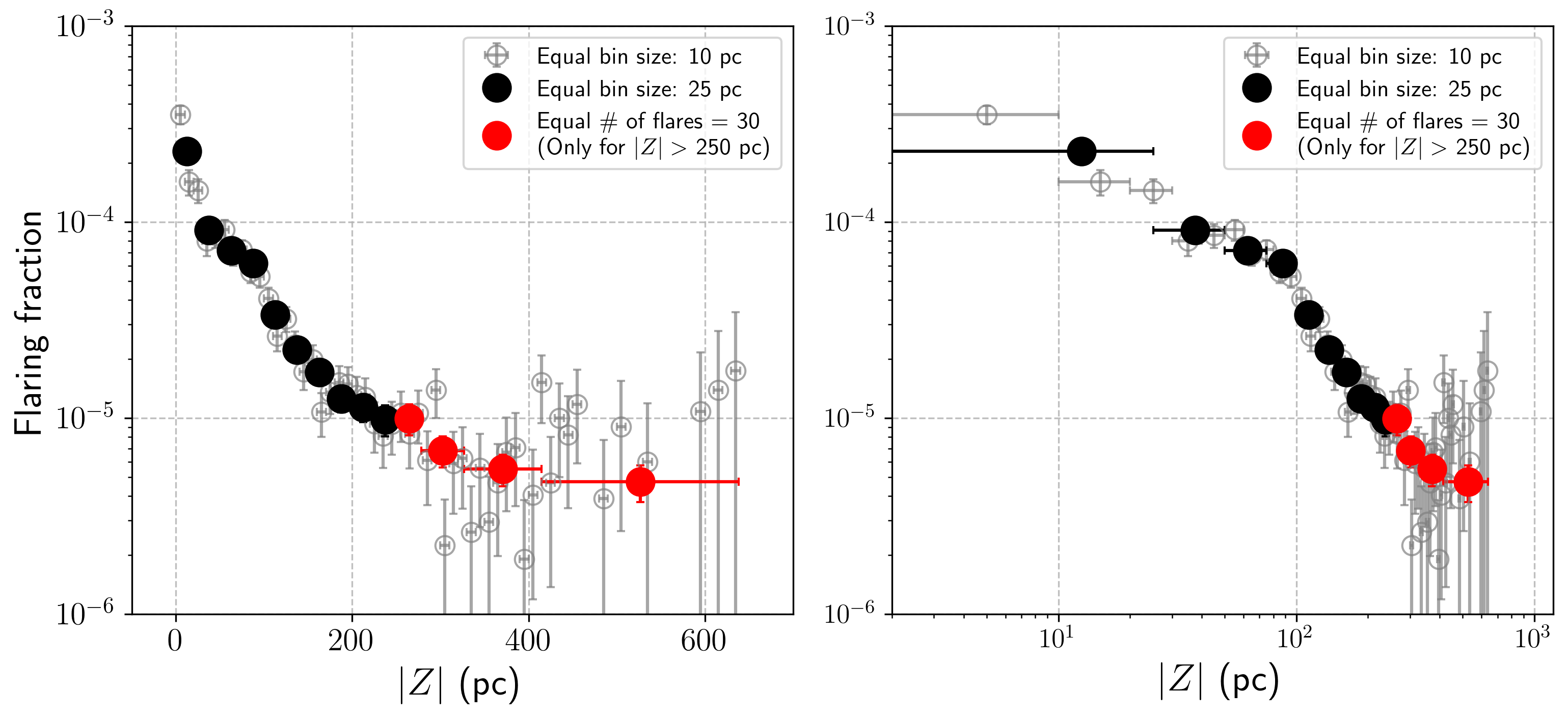}
\caption{Left: Fraction of flaring epochs as a function of vertical distance, $|Z|$, from the Galactic plane with an equal bin size of 10~pc (gray points) and the black points are those with the equal bin size of 25~pc only up to 250~pc. To increase a resolution in $|Z|$ on the distant tail over 250~pc, we adjust the width of the $|Z|$ bins that have a constant number of flare events (red points). Right: same as in the left panel but with the logarithmic scale in the x-axis.}
\label{fig:Flaring fraction vs. absolute vertical distance}
\end{center}
\end{figure*}

\begin{figure*}
\includegraphics[width=0.98\linewidth]{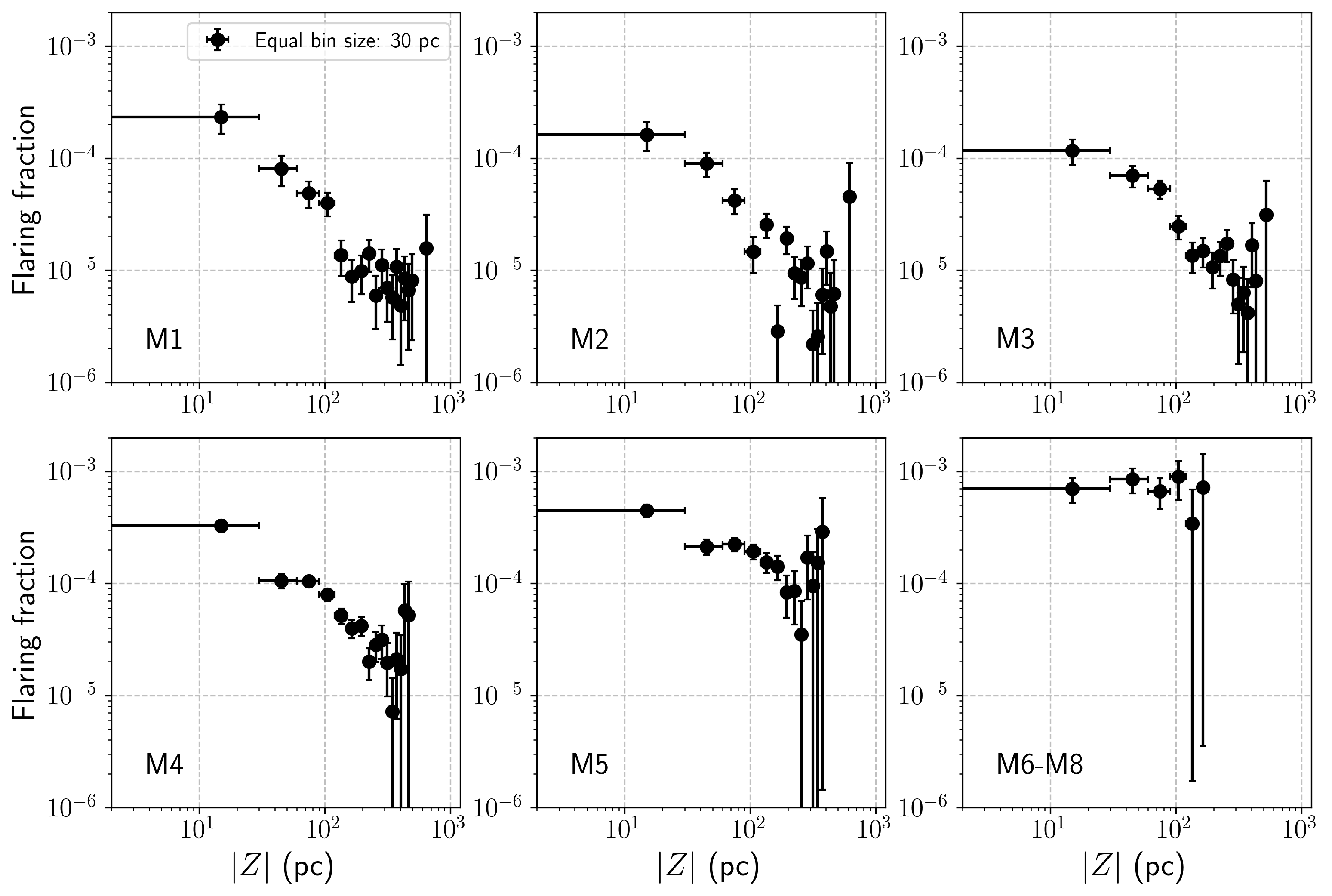}
\caption{Similar as Fig.~\ref{fig:Flaring fraction vs. absolute vertical distance} but showing the results for photometric spectral subtypes from earlier (M1: top left) to later spectral type (M6--M8: bottom right). We only show the case with the equal bin size of 30~pc for clarity.} 
\label{fig:Flaring fraction vs. Z and spectral type}
\end{figure*}

\subsubsection{Z-dependent flaring fraction}
\label{sec:Z-dependent flaring fraction}
We investigate the flaring fraction as a function of absolute vertical distance $|Z|$ from the Galactic plane (a dynamical proxy for age) for a mixture of M dwarfs of different spectral types. To get a better sense of the expected flaring fraction over distance, the results are shown with an equal bin size of 10~pc and 25~pc. To increase a resolution in $|Z|$ on the distant tail over 250~pc, we also adjust the width of the $|Z|$ bins that have a constant number of flare events. As shown in Fig.~\ref{fig:Flaring fraction vs. absolute vertical distance}, the photometric activity fractions decrease as a function of Galactic height, confirming again the consistency of our result with previous works that pointed out declining stellar activity with age \citep{West2008AJ....135..785W,West2011AJ....141...97W,Kowalski2009AJ....138..633K}. We find a {\refbf hint of a} kink in the slope of the flare fraction near 100~pc from the Galactic plane where a steep decline sets in. The appearance of the kink is not an artefact of selection effects caused by either combining of different survey components or the binning procedure. Although the number of flaring stars is not sufficient, we also see a weak hint of flattening at larger vertical distances around 400--600~pc bin. Recent work has shown a flat trend in the H$\alpha$ activity fraction in the range 400--1000~pc \citep{Zhang2021ApJS..253...19Z}.

We check how mixing spectral types may create the trend as an artefact by comparing the results on the $|Z|$ dependence by spectral subtype (Fig.~\ref{fig:Flaring fraction vs. Z and spectral type}). Flare activity decreases with $|Z|$ distance for all M dwarfs between M1 and M5, where the sample contains at least 100 flaring epochs. We still see the similar kink for spectral types M3--M5. {\refbf If this feature is real, there could be two different explanations: either it is a feature of Galactic structure, combined with a smooth temporal evolution of the flaring activity. Or it is a feature in the temporal evolution of flare rates, combined with a smooth Galactic structure. If it is a structural feature, it could affect stars of all ages, or only young and active stars. It would be interesting to check the vertical profile for stars of other spectral types as well. A mechanism for producing kinks may be a bottleneck in the dynamical heating; stars move quickly away from the Galactic plane at first but then slow down as they spend a decreasing fraction of their time in the high-density disk plane.}

The slope is useful as a point of comparison to models that attempt to understand the age-activity relation. Early-type stars have a steeper slope than the later ones. 
We attribute this to the effect of finite activity lifetimes. As discussed in previous works (e.g., \citealt{West2008AJ....135..785W, Davenport2019ApJ...871..241D}), the lower mass stars exhibit a longer lifetime of magnetic activity (or flare activity). For stars very close to the Galactic plane, there is no significant difference in the flaring fraction as a function of spectral type (see Fig~\ref{fig:Flaring fraction vs. Gaia colour}). In contrast, the difference between early- and late-type M dwarfs can be seen more clearly for those further from the plane. With the Kepler data, \citet{Ilin2021A&A...645A..42I} found an evidence that stars older than a certain age threshold show a rapid decline of flaring activity. More specifically, this transition process occurs for M1--M2 dwarfs with ages between 700~Myr and 800~Myr. For late-type M dwarfs, we expect to see the same phase transition at later stage of their lifetimes. As discussed in \citet{West2008AJ....135..785W}, the activity lifetimes of M dwarfs beyond a spectral type of M2 are typically a factor of two or more larger than earlier ones, e.g., 2.0~Gyr$\pm^{0.5}_{0.5}$ for M3, 4.5~Gyr$\pm^{0.5}_{1.0}$ for M4, and 7.0~Gyr$\pm^{0.5}_{0.5}$ for M5--M6 dwarfs. By adopting the derived ages of field stars in the local disc \citep{Casali2019A&A...629A..62C}, the median age of our flare sample is 2.63$\pm$2.35~Gyr in 7~kpc $<R<$ 9~kpc at low $|Z|<$0.5~kpc. It is thus a natural consequence that late-type M dwarfs show the shallower slopes but relatively larger magnitudes in the flaring fraction for a given $|Z|$ bin; this effect should be easier to see in the distant bins. 

One interesting feature for the coolest type (M6--M8) is that there is no decline in the flaring fraction up to 100~pc, which may be considered as another evidence of the empirical age-activity relation. Most of these stars still remained in the saturated activity regime due to their long activity lifetimes, showing little variation in upper limit of the activity levels \citep{Newton2017ApJ...834...85N,Wright2018MNRAS.479.2351W,Ilin2021A&A...645A..42I}. 
Despite differences in tracers of stellar activities, the result from our photometric tracer paints a consistent picture of the decline in stellar activity with age for different spectral types.



\subsubsection{(R, Z)-dependent flaring fraction}
\label{sec:(R, Z)-dependent flaring fraction}
We further investigate the position-dependent flaring fraction as the mean stellar age varies systematically with location $R$ and $|Z|$ across the Galactic disc. To get enough resolution on flaring activity, we split the sample into ($R-R_{\odot}, |Z|$) bins using an equal bin width for all bins (50~pc in both directions). We emphasise that the trends in the following are independent of the choice of cell size. Fig.~\ref{fig:2D flaring fraction vs. R, Z} shows the flaring fraction of M dwarfs at different two-dimensional positions within the solar cylinder, where the ranges of both directions and colour bar are all the same scale. The colour in the 2D diagram indicates the flaring fraction where redder cells are more active ones. To partly avoid possible mis-interpretation by incomplete cells (containing a small number of stars), we further split the data sets into low- ($N_\mathrm{stars} < 200$) and high-significance ($N_\mathrm{stars} \geq 200$) cells. Low-significance cells are likely to overestimate the flaring fraction (see \textquoteleft+\textquoteright~marks on the figure), mostly distributed near the edges of relatively high-significance regions (light grey). 

\begin{figure*}
\includegraphics[width=0.48\linewidth]{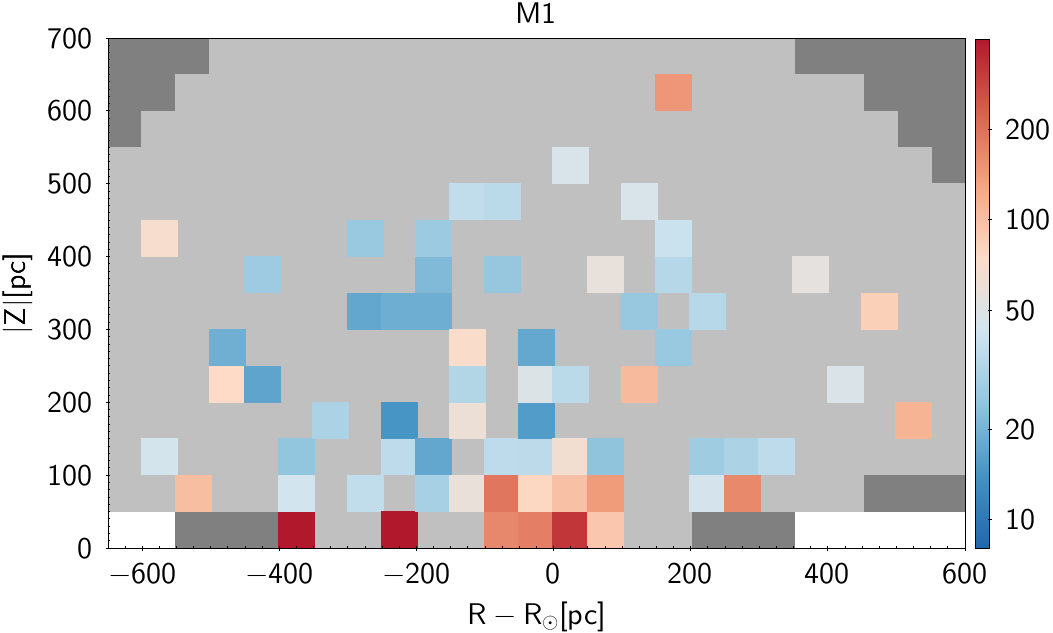}
\includegraphics[width=0.48\linewidth]{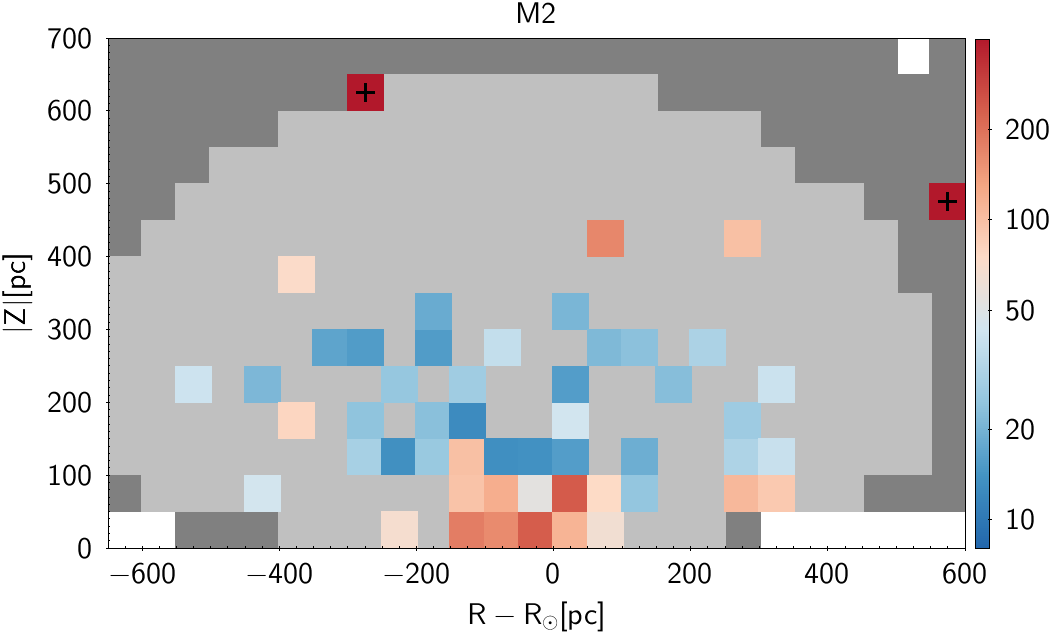}
\includegraphics[width=0.48\linewidth]{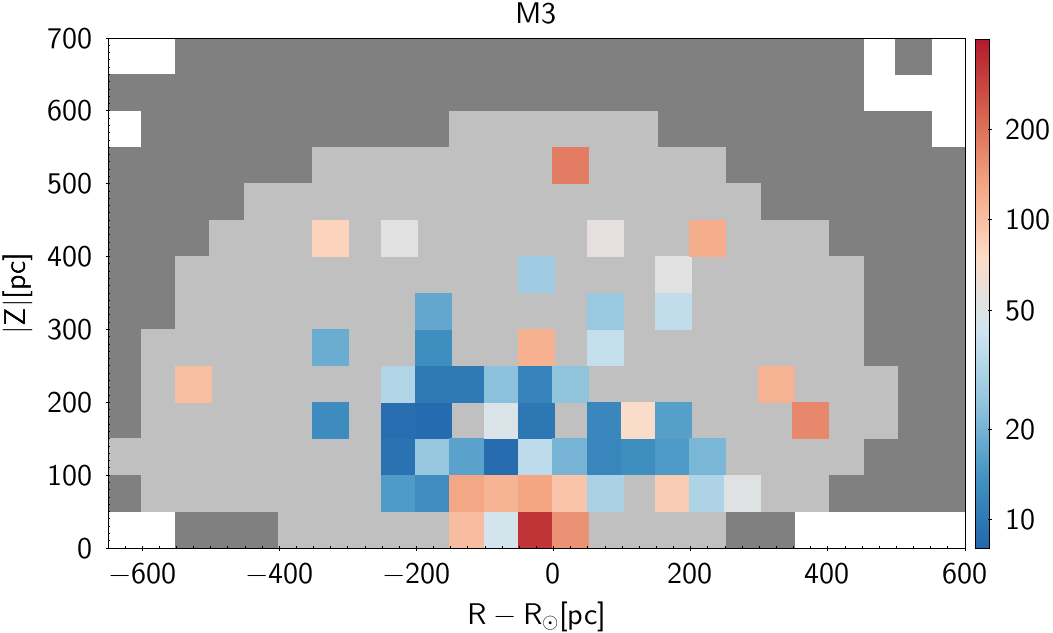}
\includegraphics[width=0.48\linewidth]{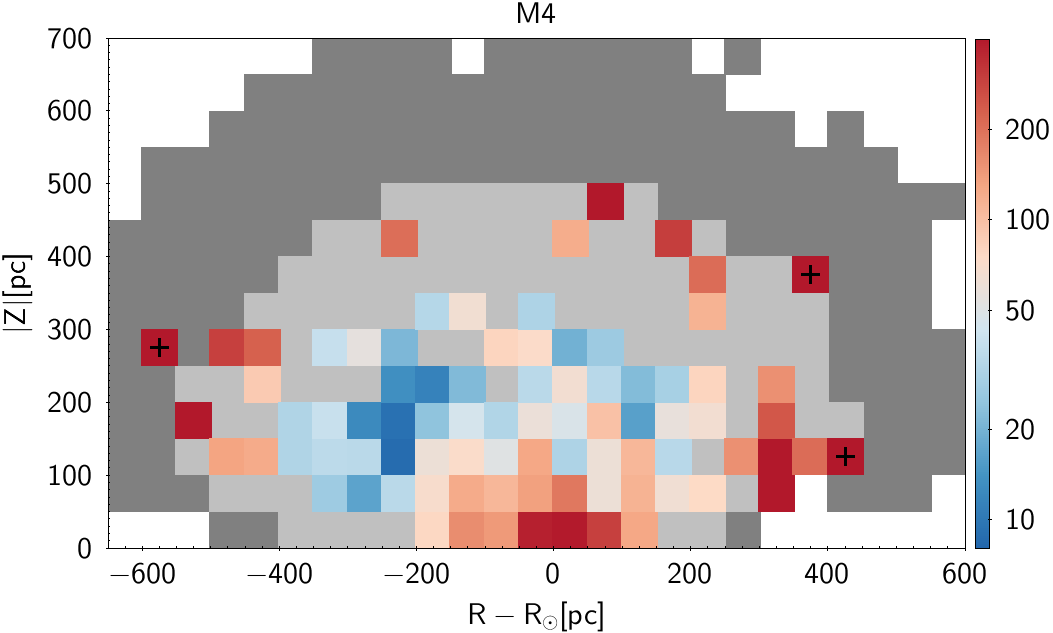}
\includegraphics[width=0.48\linewidth]{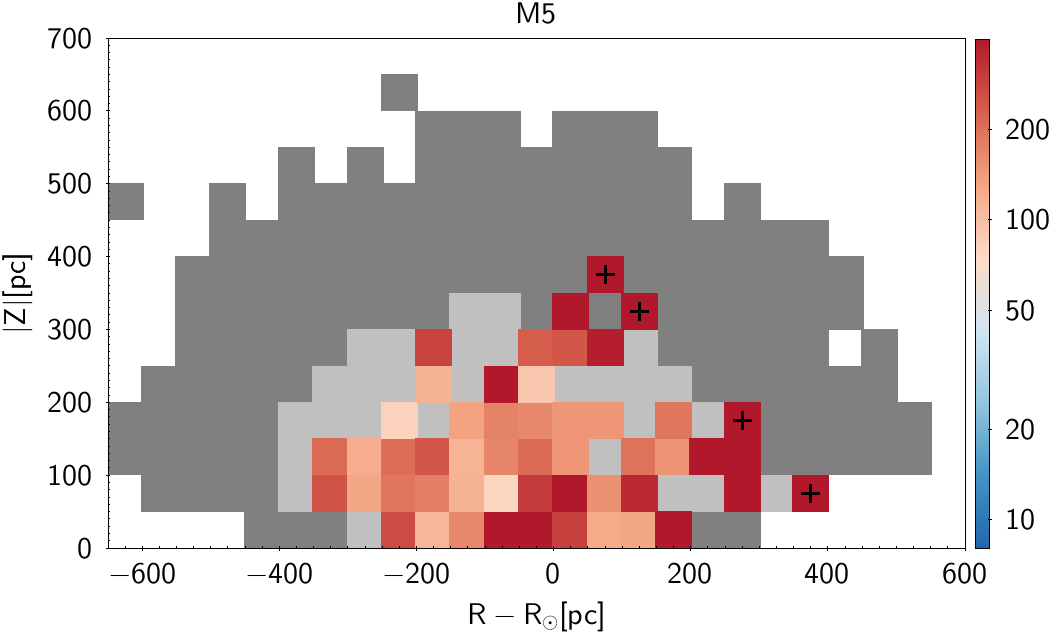}
\includegraphics[width=0.48\linewidth]{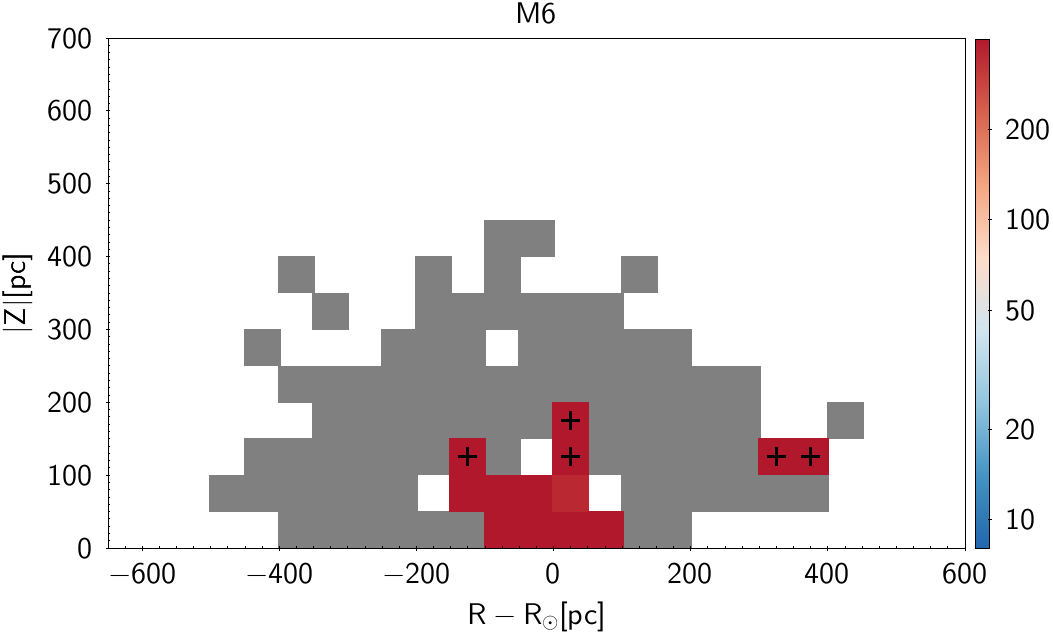}
\caption{($R-R_{\odot}, Z$) dependent flaring fraction for photometric spectral subtypes from earlier (M1: top left) to later spectral type (M6--M8: bottom right). In all panels, the size of each cell is set to 50~pc $\times$ 50~pc bins. The negative sign of $R-R_{\odot}$ axis is pointing to Galactic centre. For comparison purpose, the ranges of both axis and colour bar are all the same scale where the latter is rescaled by multiplying by the normalised constant 10$^{6}$. Redder (bluer) colours indicate more (less) active cells. To partly reduce the confusion, we overlay the \textquoteleft+\textquoteright marks onto the cells with a relatively small number of stars ($N_\mathrm{stars} < 200$). Dark and light grey cells correspond to relatively low- ($N_\mathrm{stars} < 200$) and high-significance ($N_\mathrm{stars} \geq 200$) cells with no flares, respectively.}
\label{fig:2D flaring fraction vs. R, Z}
\end{figure*}

\begin{figure}
\begin{center}
\includegraphics[width=\linewidth]{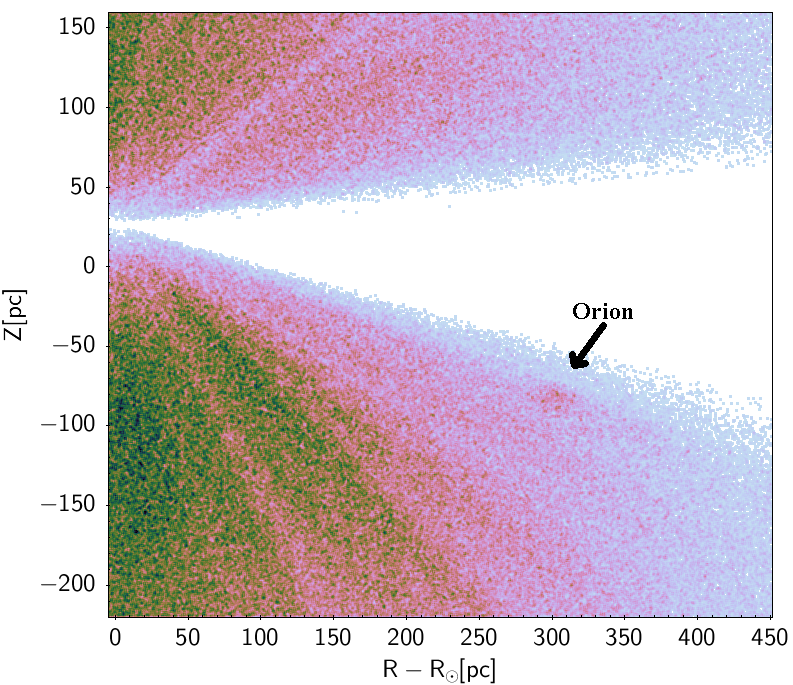}
\includegraphics[width=0.98\linewidth]{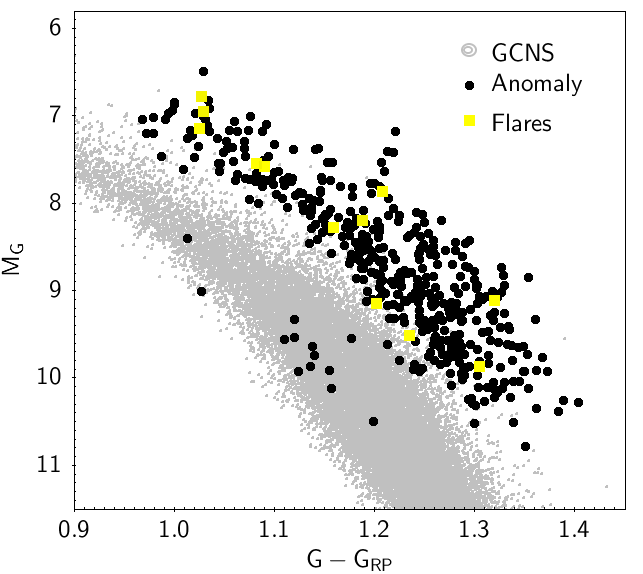}
\includegraphics[scale=0.46]{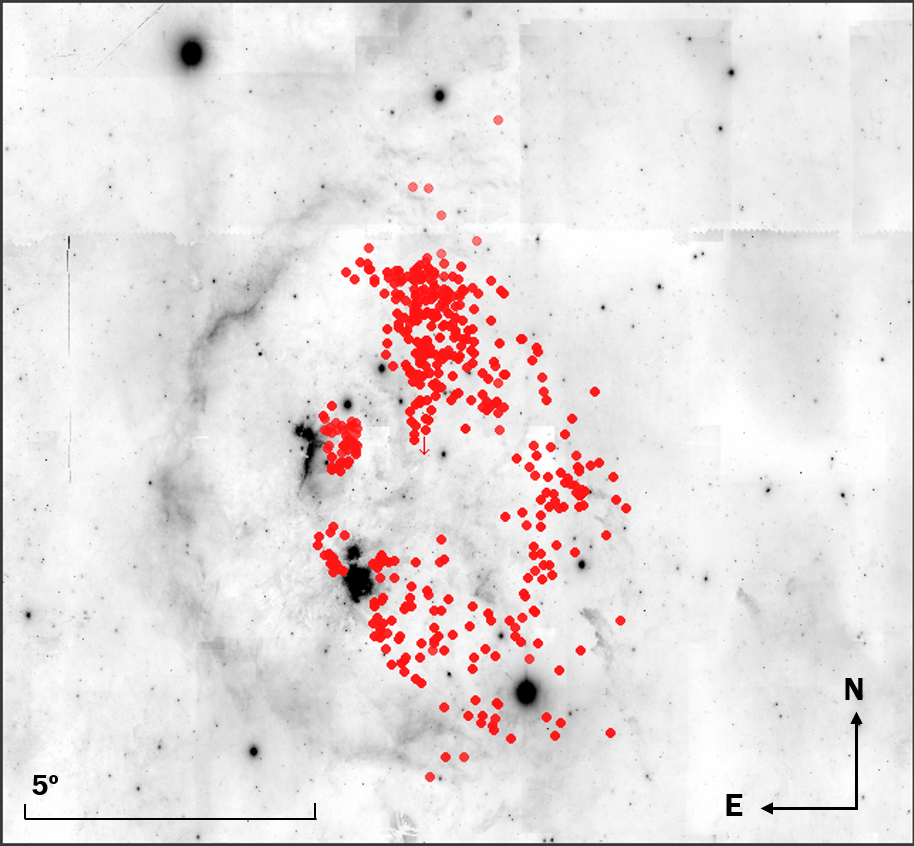}
\caption{Top: Source density distribution in the ($R-R_{\odot}, Z$) plane toward a local anomaly area around (arrow: $R-R_{\odot}$, $Z$ $\simeq$ +300~pc, -80~pc). Dark shade colour corresponds to high density regions. Middle: Colour-magnitude diagram for SMSS M dwarfs (black points) crossmatched with \citet{Kounkel2019AJ....158..122K}. Flare stars are shown as yellow squares. These are members of a nearby massive star-forming complex, Orion (Theia Group ID=13) with a mean age of about 7.85~Myr. For comparison, we show a smooth distribution of nearby M dwarfs along the main sequence in the GCNS catalogue (grey points). Bottom: Spatial position of Orion and SMSS M dwarfs (red points) on the DSS red plate image.}
\label{fig:Local anomaly}
\end{center}
\end{figure}

Our main results are as follows: 
\begin{itemize}
\item Activity gradients are predominantly seen in vertical direction for all spectral types (M1--M5) except the later one (M6--M8) due to its small extent and longer lifetime. This behaviour is similar to those reported with the H${\alpha}$ activity indicator \citep{Pineda2013AJ....146...50P, Zhang2021ApJS..253...19Z}. Although both hemispheres were sampled differently, we do not see any significant asymmetries between the north and south sides.

\item It is not easy to find any significant gradient along the radial direction in any $Z$ bins, because our flare sample only reaches a few hundred parsec from the Sun. If our flare sample extended to the outer disc of $R>$ 9~kpc ($R-R_{\odot}>$ 700~pc) where the young ($<$ 2~Gyr) population contribution is relatively high \citep{Xiang2017ApJS..232....2X}, we might expect to observe a hint of radial gradient for $|Z|<$ 1~kpc. The expected trend is that the flaring fraction increases with $R$  (i.e., negative age gradient). 

\item After excluding cells with low-significance, the results shows the unexpected high-level of activity fraction around ($R-R_{\odot}$, $Z$ $\simeq$ +300~pc, -80~pc) for nearly all spectral types. Fig.~\ref{fig:Local anomaly} shows the details of such area, displaying the source density in 2D and colour-magnitude diagram for SMSS stars cross-matched with \citet{Kounkel2019AJ....158..122K}. There are 451 clustered sources around this over-dense area, and these are members of the nearby massive Orion star-forming complex with a mean age of about 7.85~Myr. We find a mean $E(B-V)$ of $0.14\pm0.03$ for these young M dwarfs, which is lower than our sample selection criteria. After subtracting the Orion objects from the original sample, we find that all the trends discussed in previous sections (flaring fraction vs. colour and $|Z|$) remain as the contribution of Orion to the overall flare statistics is negligible. This last point emphasises that flares really are an age-sensitive activity indicator.
\end{itemize}

\section{Conclusion}
\label{sec:conclusion}
White-light flare emission is a good proxy for magnetic activity, allowing an examination of the link between the observed activity trend and age distribution. With the benefit of near-simultaneous, multi-colour observations in the SkyMapper Southern Survey DR3, the number of M dwarf flares is increased by factor of 3.5 (906 unique flare events with only $\Delta m > 0.2$ mag) compared to our previous work. Thanks to the contrast effect, flaring M dwarfs have a good visibility in the \(uvg\) bands or even in the \(ri\) bands. One important finding is that the flare-only SEDs at peak emission are intrinsically similar in nature ($T \sim$ 9\,000--10\,000 K) across stars of all spectral types. 

Based on improved distance measurements from Gaia EDR3, our sample extends to about 1~kpc from the Sun over any ($R, Z$) direction. This local sample of M dwarfs does not significantly suffer from selection effects that would alter the flare activity distribution. We present a more detailed picture of the flaring fraction in terms of spectral type and spatial location within the solar cylinder: 

\begin{itemize}
    \item We observe the flaring fraction of M dwarfs to increase from $\sim$10 to $\sim$3000 per million stars for spectral types M0 to M8. This observed trend is no different above and below the Galactic plane. For the volume-complete sample within 200~pc of the Sun (Local Bubble), we still see a similar trend of flaring fraction but the difference in rate between early- and late-type M dwarfs becomes small. When we restrict the sample to within 50~pc of the Sun, the flaring fraction becomes flat irrespective of spectral type. Thus, nearby M dwarfs, which are expected to be relatively young, have a constant fraction of flares with $>0.2$ mag contrast of 1-in-1,500.

    \item By adopting the vertical distance $|Z|$ from the Galactic disc as dynamical proxy for age, we confirm again a strong dependence of the flaring fraction on $|Z|$. 
    {\refbf We also find a hint of} a kink in the slope of the flare fraction near 100~pc from the plane where a steep decline in the overall flaring fraction is observed. We find a similar kink {\refbf among} mid-type M dwarfs (M3--M5), suggesting it is not an artefact of mixing spectral type. 
    

    \item We do not have sufficient numbers of flare-detected, distant M dwarfs to measure whether the flaring activity vs. $|Z|$ curve is flat or further declining at $|Z|>$ 0.4~kpc.
    
    \item As expected from Milky Way-like galaxy simulations and observations of evolved stars in the Galactic disc, the resulting flaring fraction in highly resolved ($R, Z$) cells shows no radial gradient at any observed $Z$ level due to the small extent of our 2D map data. If we can reach to a large distance ($|Z| <$ 1~kpc and $R >$ 9--10~kpc), we might expect to observe a negative radial gradient in age, i.e., positive radial gradient in flaring fraction).
\end{itemize}

The sample volume in this work is mainly limited by Gaia EDR3 astrometric data but also by the magnitude and random nature of flares. The full Gaia DR3 will contain sources with updated astrometric measurement and quality indicators, which is complemented with new products such as activity index and H${\alpha}$ emission from the BP/RP spectra. Combining our photometric activity indicator with measurements of different indicators will give further insight into age-dependent magnetic activity of M dwarfs in the Galactic disc.

\section*{Acknowledgements}
{\refbf We are grateful to an anonymous referee for their comments and ideas that improved the manuscript.} SWC acknowledges support from the National Research Foundation of Korea (NRF) grant, No. 2020R1A2C3011091, funded by the Korea government (MSIT). CAO was supported by Australian Research Council (ARC) Discovery Project DP190100252. The national facility capability for SkyMapper has been funded through ARC LIEF grant LE130100104 from the Australian Research Council, awarded to the University of Sydney, the Australian National University, Swinburne University of Technology, the University of Queensland, the University of Western Australia, the University of Melbourne, Curtin University of Technology, Monash University and the Australian Astronomical Observatory. SkyMapper is owned and operated by The Australian National University's Research School of Astronomy and Astrophysics. The survey data were processed and provided by the SkyMapper Team at ANU. The SkyMapper node of the All-Sky Virtual Observatory (ASVO) is hosted at the National Computational Infrastructure (NCI). Development and support the SkyMapper node of the ASVO has been funded in part by Astronomy Australia Limited (AAL) and the Australian Government through the Commonwealth's Education Investment Fund (EIF) and National Collaborative Research Infrastructure Strategy (NCRIS), particularly the National eResearch Collaboration Tools and Resources (NeCTAR) and the Australian National Data Service Projects (ANDS). This work has made use of data from the European Space Agency (ESA) mission {\it Gaia} (\url{https://www.cosmos.esa.int/gaia}), processed by the {\it Gaia} Data Processing and Analysis Consortium (DPAC, \url{https://www.cosmos.esa.int/web/gaia/dpac/consortium}). Funding for the DPAC has been provided by national institutions, in particular the institutions participating in the {\it Gaia} Multilateral Agreement. 

\section*{Data availability}
The data underlying this article were accessed from the SkyMapper website  \url{http://skymapper.anu.edu.au}. The derived data generated in this research will be shared on reasonable request to the corresponding author.



\bibliographystyle{mnras}
\bibliography{WOMBAT_DR3}

\end{document}